# AI-Driven Transformation of Platform Work: A Task-Based Model for Measuring Labour Disruption and Vulnerability, with Early Evidence from Online Freelancing in Kenya

Farhan Latif
United Nations University, Macau
University of Waterloo, Canada
latif@unu.edu

## Abstract

Assessments of AI's impact on labour often rely on conventional occupational categories or broad market indicators. This dominant trend, however, obscures how tasks within the same occupation vary in practice across work arrangements. This limitation is particularly important in online freelancing, where workers are typically remote and devote substantial time performing unpaid tasks. This article introduces a task-based framework that distinguishes AI-driven disruption, defined as the extent of change in how work is performed, from livelihood vulnerability associated with that change. The framework assesses the potential for automation, substitution, and enhancement across four categories of platform work: writing, virtual assistance, transcription, and web and software development. The model estimates high disruption across all four categories, while livelihood vulnerability scores are lower and vary by occupation. Potential efficiency gains on unpaid tasks partly offset potential wage impact associated with paid work in the model. The article compares these estimates with evidence from a survey of 52 online freelancers and 20 follow-up interviews in Nairobi, Kenya. Workers report changes in how they perform tasks and intensifying competition, while their accounts show how adapting their skills and using AI strategically shape their concerns about job loss. The findings show how task-based modelling can provide a baseline for assessing AI-driven changes in platform work and how workers' experiences differ from projections and help explain how these changes may threaten their livelihoods.

**Keywords**: AI and Work; Platform Work; Task-Based Assessment; Livelihood vulnerability; Unpaid platform labour

## *Introduction*

The evolution and ongoing diffusion of artificial intelligence (AI) tools hold significant potential to reshape work and worker productivity. From a livelihood sustainability perspective, this transformation of work is associated with significant uncertainty surrounding its effects on job availability, skills demand, and wages (Agrawal et al., 2019). As general-purpose AI tools continue to evolve and grow in adoption, their impact on labour markets demands deeper, evidence-based analysis. This paper analyzes the complex impact of digitalization and artificial intelligence (AI) on the livelihood sustainability of Kenyan freelance platform workers.

The practical impact of this technological transition on workers' livelihoods remains unclear without empirical investigation (Hui et al., 2023; Liu et al., 2023). This uncertainty arises from several research gaps and challenges. There are significant differences between theoretical AI exposure and its practical impact on jobs (Albanesi et al., 2023; Gmyrek et al., 2023). The adoption of AI tools across firms, regions, and individual workers remains uneven, while the capabilities of these tools are constantly evolving and advancing with each update (E. W. Felten et al., 2023). More importantly, these tools have a multifaceted impact on workers' livelihoods, as they simultaneously function to substitute and complement work tasks (Agrawal et al., 2019).

It is likely that AI-driven automation may depress wages and reduce job quality by narrowing the range of tasks where human expertise remains indispensable (Occhipinti et al., 2024). However, historically, theoretical models predicting technology's impact, including widespread labour market disruption and job losses, often remain disconnected from the actual effects on the labour market (Agrawal et al., 2019). Yet, competing perspectives also argue that this time around, the disruptive impact on the economy and labour market would be different since the ability of AI tools to handle non-routine cognitive tasks offers the type of productivity growth and labour market disruption that may not self-correct, leading to labour redundancy and broader recessionary pressure (Occhipinti et al., 2024).

Recognizing the impact of this technological change on occupations and employment, the recent discourse on AI's potential impact on jobs (Acemoglu et al., 2022; George et al., 2023; Gmyrek et al., 2023) has primarily focused on exploring the risks and impacts within conventionally recognized occupational categories and job titles. While building a macro-level understanding of potential AI risks and opportunities, this broad perspective often overlooks the nuanced ways in which work is organized (e.g., platform work) within the same occupations and industries, potentially leading to varied AI effects on skills and tasks within a given occupation.

While AI tools could alter work processes and productivity metrics, the platformization of knowledge work [1] fundamentally transforms how labour is organized as platforms deconstruct traditional jobs into task-based units, challenging labour norms and traditional occupational categories (De Stefano, 2015; Mark Graham et al., 2017). I build on the understanding that even when job and occupation titles stay the same, the nature and frequency of tasks, methods, and tools used can differ significantly among workers, depending on how work is organized. These variations suggest that AI technologies may differentially impact worker groups, reshaping the demand for specific skills and the remunerative potential of tasks while leaving others largely unchanged. As a result, a more granular, task-level analysis is essential to accurately assess AI's true impact on workers within an occupation.

This need for detailed examination is particularly acute in the domain of knowledge work, where precarious working conditions are increasingly evident, such as freelancing through digital platforms. The growth of online labour platforms has transformed the employment landscape, especially in developing regions where education gains and digital connectivity have significantly improved, but decent work opportunities are relatively scarce (Bonina et al., 2021). Freelancers operating on platforms such as Upwork and Fiverr engage in a variety of tasks across different occupations, requiring a range of digital skills, making them uniquely susceptible to the nuanced effects of AI on specific tasks and work availability.

Although these freelancers in developing regions have historically leveraged competitive pricing to secure offshored jobs (Mark Graham et al., 2017), the growth of AI systems capable of handling tasks at scale could threaten to disrupt this advantage, potentially reshaping global demand for and pricing of human labour. The precarious nature of platform work and the lack of employment safety nets could further amplify these vulnerabilities, making this group of the workforce particularly exposed to rapid technological shifts and global market fluctuations (Standing, 2019). Despite extensive discussions on AI's potential impact on employment, there is a lack of empirical research that systematically assesses this disruption and how AI affects specific tasks within various freelance occupations, especially in developing regions (Hui et al., 2023). Understanding which tasks are most susceptible to AI disruption and recognizing the nature of this disruption is essential for freelancers, policymakers, and platform operators to develop strategies that mitigate risks and leverage AI for sustainable employment.

In this research, the notion of generative artificial intelligence (AI) refers to computational systems designed as interactive products capable of generating seemingly new, meaningful content, such as text and images, from training data (Brown et al., 2020). These products are built on recent breakthroughs in machine learning that have paved the way for large language models (LLMs) that harness vast datasets to produce human-like text and perform sophisticated language and data tasks (Brynjolfsson et al., 2023; Sutton & Barto, 2018). A generative AI model can be either general-purpose or specialized (i.e., industry-specific), built on a machine-learning architecture that is designed to create content by identifying and replicating complex patterns in data (Acemoglu & Restrepo, 2019).

However, the automation aspect of AI also encompasses AI agents. These somewhat autonomous software entities recognize the common digital environments, such as using a web browser to perform certain tasks, make decisions, and execute actions to achieve specific goals based on given objectives, which could contribute to the automation of common workplace tasks (Kolt, 2025). It represents a different level of automation, as traditional automation primarily involved programming systems to perform repetitive, rule-based tasks, thereby optimizing operational efficiency (Acemoglu & Restrepo, 2020; Brynjolfsson & McAfee, 2014). By contrast, the newer conceptualization of AI is characterized by adaptive learning capabilities that extend its influence beyond synthetic content creation and routine task execution (Brown et al., 2020; Kolt, 2025).

In this paper, I focus on how AI's capabilities in enhancing labour productivity, automating tasks, and substituting human labour pose both opportunities and challenges for freelance platform workers who rely on digital platforms for employment. I assess the extent to which this technological shift reshapes work and whether this change results in livelihood vulnerabilities for those working in platform-dependent jobs. The occupations examined in this analysis include freelance writing, web and software development, virtual assistance, transcription, and freelance transcription services.

[1] The notion of Knowledge work here refers to labour that primarily involves a higher level of cognitive complexity and specialized expertise, rather than manual or routine tasks. It is commonly defined as work in which knowledge serves as both the principal input and output (Allison Woodruff et al., 2023; Alvesson, 2001). Within the platform economy literature, this distinction is particularly salient, as it helps differentiate cognitively demanding and interactive gigs, such as writing, software development, graphic design, and customer service, from highly repetitive or routine tasks such as image tagging, clickwork, as well as manual services such as ride-hailing and delivery work (Huws, 2014; Kalleberg & Dunn, 2016).

I take the approach to structure this inquiry around two interrelated yet distinct dimensions of AI's impact on freelance work: disruption and vulnerability. In digital labour markets, disruption and vulnerability are not incidental by-products but embedded within the design of platform work structures. Disruption operates on two levels. First, it fragments work into decontextualized, modular tasks that decouple execution from decision-making, flatten occupational hierarchies, and enable new forms of remote oversight and productivity tracking (Gray & Siddharth, 2019; Kellogg et al., 2020). And second, it restructures labour relations through algorithmic control, flexible contracting, and the externalization of risk (Graham et al., 2019; Srnicek, 2017). AI, however, may introduce additional layers of change and reshape the interaction between workers and the worker providers. This duality makes it challenging to differentiate the incremental impact of AI from the systemic instability already present in the platform economy.

In the context of this research, disruption refers to any drastic departure from the established status quo regarding both the nature of work and its potential for remuneration. It denotes the overall transformation in how tasks are configured and executed, encompassing both opportunities (e.g., enhanced productivity, new work modalities, the need for reskilling) and challenges (e.g., task reorganization, shifting skill requirements) (Autor, 2015; Brynjolfsson et al., 2018; Huang & Rust, 2021). From a platform work perspective, disruption is not merely a technical adjustment but a continuous restructuring of work, one that obliges platform workers to navigate new regimes of control, competition and compensation and impose significant transition costs (De Stefano, 2015; Wood et al., 2019). This deskilling and reskilling due to continuous reorganization of work remains a hallmark of platform work (Latif, 2024; Srnicek, 2017).

This conceptualization of platform work disruption builds on the understanding that disruption is a multifaceted process that involves significant, often rapid, reconfigurations of established practices, processes, and structures within workplaces and industries (Christensen, 1997; De Stefano, 2015). Significant changes in work practices, whether introduced through new tools and technologies such as labour substitution via generative AI or organizational arrangements like the platformization of work, are not neutral innovations but deliberate shifts that unsettle established labour processes. While such a form of disruption is not inherently harmful and can open up opportunities for worker groups to reposition themselves within emerging opportunities, it nonetheless demands a period of adjustment that may precipitate struggles at the worker level.

Vulnerability, on the other hand, refers to the structural exposure of workers to factors that collectively reduce their livelihood opportunities by causing wage cuts and job losses (Cingolani, 2021). The conceptualization of vulnerability focuses specifically on the adverse economic outcomes that may arise from these changes, such as job loss, wage reduction, or the sudden devaluation of skill capital (De Stefano, 2015; Standing, 2019). Unlike disruption, which broadly denotes the reconfiguration of work practices, vulnerability in this context emphasizes the economic risks and livelihood insecurities that arise when new forms of production render work opportunities inaccessible or less economically competitive.

This paper quantifies disruption and vulnerability in the platform work economy. The disruption score in this study quantifies the extent to which AI is likely to alter a task, reflecting the overall magnitude of change in how work is performed. This metric is crucial because it highlights the need for reskilling as the retooling of tasks potentially demands retraining, regardless of whether the impact is perceived as positive or negative. The focus on disruption underscores that AI will bring about significant shifts in task execution, necessitating greater adaptability among workers. The vulnerability score assesses how susceptible a task is to being affected by AI in ways that directly impact workers' access to work, particularly for those in the Global South. This metric is important because it sheds light on the potential risks to income and employment stability for freelancers who may already be in precarious economic situations.

I examine the potential likelihood of disruption and vulnerability within the selected platform work categories using a task-based analysis model. By adopting a task-based analysis of AI impact, which provides a framework for isolating AI's specific effects from the broader disruptions inherent to other aspects of platform-mediated work, this model disentangles these intertwined influences, thereby offering a more precise understanding of how AI contributes to both the disruption and vulnerability of platform workers. The following key questions guide this research.

RQ1: How can we systematically measure the magnitude of AI-related occupational disruption and the extent of worker vulnerability across different platform work occupations?

RQ2: To what extent do AI's task-enhancing capabilities mitigate its labour substitution effects across different occupational categories within platform work?

RQ3: Which core tasks within selected platform work sub-categories drive disruption and vulnerability, and how does this susceptibility vary across occupational categories?

RQ4: How well do such disruption and vulnerability projections correspond with workers' experiences and perceptions?

The first section addresses question 1 by building on existing research to develop and validate quantitative measures that capture both the overall disruption and the vulnerability induced by AI across different freelance occupations. Existing measurement techniques that rely solely on platform labour market statistics often fall short of capturing these nuanced dynamics (see, e.g., Gmyrek et al., 2023; Jing Liu et al., 2023), highlighting the need for more refined metrics (discussed further in the literature review section). I introduce a measurement model that builds on the nature of tasks, such as cognitive routine and non-routine tasks (Autor, 2015; Frey & Osborne, 2017) and the potential impact of AI, such as substitution, automation risk, and worker performance enhancement (Eloundou et al., 2023; Mäkelä & Stephany, 2024).

The second section operationalizes this model to answer questions 2 and 3. The second question recognizes the dual nature of AI: its capacity to both substitute for and enhance human labour, potentially mitigating the adverse effects on vulnerable workers. This question aims to assess the extent to which AI's task-enhancing capabilities can mitigate its displacement effects on the selected occupational categories. The third question examines a range of core tasks within selected freelance occupations to identify the drivers of both disruption and vulnerability across occupational categories. A granular understanding of these drivers is critical for explaining why freelancers may shift between occupational categories or experience divergent outcomes, and it offers valuable insights for freelancers, skill trainers, and policymakers alike.

The third section examines the empirical validity of these assumptions by analyzing both qualitative (n = 20) and quantitative data from my field study, which involved 52 platform workers in Nairobi, Kenya. This section is a direct response to question 4 and addresses the empirical gap as highlighted by scholars (Hui et al., 2023; Liu et al., 2023). It examines how well these disruption and vulnerability metrics correspond with freelancers' experiences and economic outcomes on digital platforms. By linking task-level assessments to workers' direct perspectives, I evaluate the external validity of these measures and illuminate their lived experiences.

This study contributes to multiple bodies of literature by advancing the understanding of AI's impact on digital platform work at both task and occupation levels. First, it extends the literature on the task-based measurement methods to capture the magnitude of the AI-driven labour market transition by providing empirical data and a methodology identifying tasks vulnerable to AI disruption (Acemoglu et al., 2022; Agrawal et al., 2019; Autor, 2015). Second, it builds on the scholarship on digital platform labour markets and the impact of AI on platform-mediated knowledge work (Mark Graham et al., 2017; Hui et al., 2023; Liu et al., 2023; Melia, 2020) by analyzing how AI reshapes tasks that form the livelihood of freelance platform workers. Third, by highlighting both the challenges and opportunities AI presents, ranging from the need to upskill workers and redesign tasks to the potential for enhanced productivity, this research informs workers and development agents seeking to foster resilience in the digital economy. The study uses the terms freelancers, gig workers, and platform workers interchangeably to denote the global workforce of knowledge workers who leverage digital platforms to find gig work that is typically done remotely.

## *Literature Review*

Recent literature highlights a growing interest in measuring the early-stage effects of generative artificial intelligence in reshaping the nature of work, e.g., Acemoglu & Restrepo, 2019; Felten, Raj, Manav, et al., 2023; Frey & Osborne, 2017; Gmyrek et al., 2023. Current research has generally followed two methodological streams.

### Methodological Approaches and Perspectives in AI and Labour Research

On the one hand, some studies analyze job market data to track changes in skill demand and labour supply over time. e.g., (Acemoglu & Restrepo, 2019; Hui et al., 2023). These studies adopt an approach that emphasizes the use of observable market-level trends and dynamic changes in labour demand to understand the broad influence of AI tools.[2]

On the other hand, another set of studies takes a task-based perspective, focusing on the underlying occupational structures (Eloundou et al., 2023; E. W. Felten et al., 2023; Frey & Osborne, 2017; Liu et al., 2023). These bodies of literature examine the broader impacts of AI on occupations and labour markets, concentrating on stable, well-defined roles and skills (Frey & Osborne, 2017; Webb, 2019). For example, the International Labour Organization's report on AI's potential effects on jobs (Gmyrek et al., 2023) offers a global analysis of AI's potential impact on jobs. Following the techniques of Eloundou et al. (2023), their method builds on sequential GPT-4 API calls to generate task-level automation scores aggregated at the occupation level. Their findings highlight, for instance, that clerical support occupations exhibit high automation potential, while other occupational groups show much lower exposure.

Frey & Osborne (2017) and Felten, Raj, Manav, et al. (2023) both draw on standardized O*NET[3] data to assess the susceptibility of occupations to AI automation, yet they employ distinct methodological approaches. Frey and Osborne employ a Gaussian process classifier to estimate the probability of computerization across 702 detailed occupations, examining task characteristics to distinguish between routine and non-routine tasks, as well as manual and cognitive tasks. Their analysis suggests that jobs requiring creative or social intelligence appear more resilient.

[3] ONET (Occupational Information Network) is an online database sponsored by the U.S. Department of Labour that offers detailed,

In contrast, Felten, Raj, Manav, et al. (2023) adapt the AI Occupational Exposure (AIOE) framework to capture the impact of recent advances in large language models (LLMs), such as those embodied by different versions of ChatGPT3. By assigning weights to the AI models and human skills within their exposure matrix, they reveal that specific education-related roles and digital work, such as telemarketing, are notably vulnerable. However, they find a positive correlation between higher wages and exposure to these technologies. Although both studies leverage ONET data, Frey and Osborne offer a broad, generalized risk assessment. In contrast, Felten et al. provide a technology-specific perspective highlighting how task reconfiguration within particular domains may evolve. Notably, neither study captures the contextual occupational variability, such as differences between traditional and remote or freelance work, pointing to a gap that future research should address.

While these approaches towards the task-based assessment of AI impact on work offers a deeper understanding of the intrinsic automation potential in various occupational structures, it is important to also consider the differences that emerge when work is organized differently. A traditional understanding of occupational structures differs from, and may sometimes overlook, the operational nuances in digital work contexts (Wood et al., 2019). Hence, I contend that by aggregating data at the occupational level, the analysis, when applied to platform work, may overlook critical nuances within digital work contexts or the enhancement impact of the same technology.

## Existing Literature on the Impact of AI on Platform Work

Among the relevant quantitative research studies covering digital platforms, Jing Liu et al. (2023) utilized a dataset of jobs, bids, transactions, and freelancer profiles from Upwork to analyze how generative AI reshapes labour demand and supply. They created submarkets (in lieu of occupations) based on skill tags and applied difference-in-differences and triple-differences methods to capture the impact of ChatGPT's availability on these submarkets. Their findings reveal a significant drop in labour demand (fewer jobs posted and fewer participating clients) for text-based and coding-related submarkets and fewer new freelancers entering them. Jing Liu et al. (2023) note that many text-oriented freelancers also begin bidding on programming tasks, citing ChatGPT's ability to lower coding barriers. Notably, their analysis shows no such effect for image-based work. According to Jing Liu et al. (2023), this is presumably because current image-generation tools remain narrow in scope. These results highlight the fluid and uncertain boundaries of general-purpose AI, underscoring freelancers' adaptability through skill transitions in response to rapid technological change.

While Liu et al. (2023) provide valuable market-level evidence on how ChatGPT reduces demand for certain skill-based submarkets and reshapes freelancer bidding, their analysis remains at a relatively aggregated scale, focusing on broad submarkets (e.g., text versus programming) rather than the specific tasks that vary within an occupation. Moreover, they do not employ dedicated metrics to pinpoint the magnitude of disruption or worker vulnerability arising from AI adoption. This gap is precisely where the present research intervenes, conducting a task-level inquiry into freelance work to capture how AI reshapes both the nature of tasks and workers' livelihood security.

Among the studies focusing on job availability data, Xiang Hui et al. (2023) using Upwork platform job availability data shows that after ChatGPT's release, freelancers offering writing-related services experienced a decline in work. They highlight that among those freelancers who receive work, the study documents a 4.7% decrease in the number of jobs and a 5.1% reduction in monthly income. Notably, their results suggest that higher-quality freelancers are less adversely affected by these technological shifts. In a complementary analysis, they extend their approach to image-generation technologies (including DALL-E 2 and Midjourney) and find that freelancers in design, image editing, and art services experience similarly negative impacts on job counts and compensation. It is worth noting that these findings related to graphic work contradict the findings by Jing Liu et al. (2023) that estimated lower disruption among image-related work, demonstrating the limits and complexity of measuring impact on freelance work on a granular level.

## Macro-Level and Theoretical Impact of AI on Work

Broadly, a central question emerging from the literature involves whether AI tools merely boost productivity or substantially displace human labour (Acemoglu & Restrepo, 2019; Agrawal et al., 2019; Brynjolfsson et al., 2018). The outcome can vary significantly across different occupations and tasks depending on how the technology is integrated, whether as an assistant or a near-total replacement. Key scholars in this field argue that large language models can perform a wide array of tasks at or near human-level quality, suggesting they can serve as both substitutes for and

---

standardized information on job requirements, worker attributes, and occupational characteristics. (*O*NET Resource Center*, n.d.)

complements to human labour (Agrawal et al., 2019; Eloundou et al., 2023; Mäkelä & Stephany, 2024).

On the one hand, some studies suggest that AI can function as a powerful complement to human labour rather than complete automation or substitution (Agrawal et al., 2019). By automating repetitive or lower-value tasks, AI frees individuals to devote more time and cognitive resources to creative, strategic, or social dimensions of their work, areas in which human judgment remains indispensable (Huang & Rust, 2021; Mäkelä & Stephany, 2024). On the other hand, a competing view warns that dependence on AI may erode essential worker capabilities, a phenomenon referred to as skill atrophy, where individuals may become less adept at certain tasks when they overly rely on algorithms and automated systems (E. Felten et al., 2023).

Increasingly, however, the scope of AI raises fresh concerns. According to (Eloundou et al., 2023), large language models can handle both routine and some non-routine cognitive tasks[4], challenging the notion that labour markets will naturally absorb disruptions by creating entirely new, offsetting roles. This creates a paradox where rapid productivity gains, if not matched by the development of new tasks requiring human intervention, could intensify labour-market frictions and potentially create recessionary pressures (Acemoglu & Restrepo, 2020; Brynjolfsson et al., 2021). Still, the ability to automate a task does not automatically translate to immediate or universal adoption, as cultural, economic, and policy factors often mediate its real-world impact (Acemoglu, 2021). Although this macro-level AI impact has significant policy value, the realities of individual workers facing displacement and deskilling and their reskilling struggles are context-specific and require a microlens to examine this technological disruption.

### Limitations of Traditional Occupational Aggregation in Digital Platform Work

This emphasis on traditional employment categories can obscure the intricacies of freelance or gig-based work, where individuals frequently perform a significantly different set of tasks, sometimes under the same occupation label, as discussed in the existing literature on the occupational impact of AI (Tolan et al., 2019). The traditional occupation-centric lens has been critiqued for its limited applicability to fragmented and diverse platform work, where boundaries between these categories can blur because many online workers take on diverse tasks that do not always fit neatly into a single traditional occupation (Kässi & Lehdonvirta, 2018; Wood et al., 2019).

These scholars argue that standardized occupational titles, while useful for large-scale labour market analyses, may poorly capture the granular task mix that defines freelancing. Consequently, the theoretical predictions derived from mapping traditional occupations to AI exposure may not align with the actual work experiences of freelancers, particularly in low- and middle-income countries where digital platform labour is rapidly expanding. This misalignment underscores the necessity of applying a more dimensional approach centred on freelancing-specific tasks and worker experiences rather than relying solely on static occupational labels.

In platform-based freelancing, questions of AI exposure and displacement risk are particularly complex and multifaceted. Even workers formally grouped under the same occupational title (for instance, freelance writer) may carry out a portfolio of tasks such as drafting marketing copy, researching content, editing standardized documents, producing keyword-rich web content, administrative work, and repetitively applying for jobs (Kässi & Lehdonvirta, 2018). Each task presents its own degree of complexity, creativity, and susceptibility to the impact of AI. The occupational exposure index by Felten, Raj, et al. (2023) provides a starting point, but such an approach risks misinterpretation if applied to freelance work through digital platforms, where activities often diverge widely from those of traditional employees with the same job title.

These perspectives suggest that the real-world impact of generative AI on freelance occupations, particularly in the Global South, is influenced by an interplay of economic, social, and technological factors. Critically, there is a need to measure not only the potential for displacement but also the degree to which AI tools might augment and diversify existing worker capacities (Brynjolfsson et al., 2021; Eloundou et al., 2023).

## *Conceptual and Analytical Framework for Task-Based AI Impact Measurement*

I build on the understanding that the collection of platform work tasks within an occupation category is similar but not the same as the tasks within a traditional occupation category. Therefore, grasping the layered reality of platform workers requires going beyond conventional occupational tasks, focusing instead on the distinct tasks that define the work a platform worker performs. A task-level assessment model of AI impact, therefore, offers a granular gauge of how AI might disrupt or augment platform-mediated work while accounting for the proportion of time or weight each subtask represents, acknowledging that a single occupation might include both easily automated duties (e.g., routine data entry) and highly creative functions (e.g., brand storytelling) (Frey & Osborne, 2017). Simultaneously, the early stage of this research topic underscores the need to align empirical investigation with theoretical models that are flexible enough to accommodate AI's rapidly evolving capabilities.

[4] Routine tasks are activities that involve repetitive, predictable, and well-structured processes, often requiring little to no creativity or complex decision-making. Non-routine tasks, on the other hand, involve problem-solving, critical thinking, or creative decision-making and are less predictable. (Autor, Levy, & Murnane, 2003).

The suggested framework below (Figure 5.1) quantifies the likelihood of how individual tasks within a particular category of platform work are being transformed by AI, capturing the nuanced interplay among automation, substitution, and enhancement impacts. It implies that an objective analysis of the characteristics of tasks within an occupation can clarify the nature of AI's impact on those tasks. Additionally, determining the relative significance of each task and accounting for often-overlooked unpaid tasks can ensure a more balanced understanding of this impact at the occupational level, helping to recognize the magnitude of disruption and potential livelihood vulnerability.

Rather than treating the impact dimensions as binary outcomes, the model calculates probabilistic scores for work tasks, reflecting that slight variations in task characteristics can shift the balance between being entirely automated, substituting the majority of human labour, or only benefiting the worker but enhancing their abilities. This framework provides a tool to generate robust occupation-level disruption and vulnerability scores by aggregating these task-level probabilities through task weightings that account for both economic significance and unpaid labour. In doing so, it identifies which tasks and occupations are most likely susceptible to AI-driven change and reveals the underlying trade-offs, such as the constrained potential for worker enhancement when a task already exhibits high automation characteristics

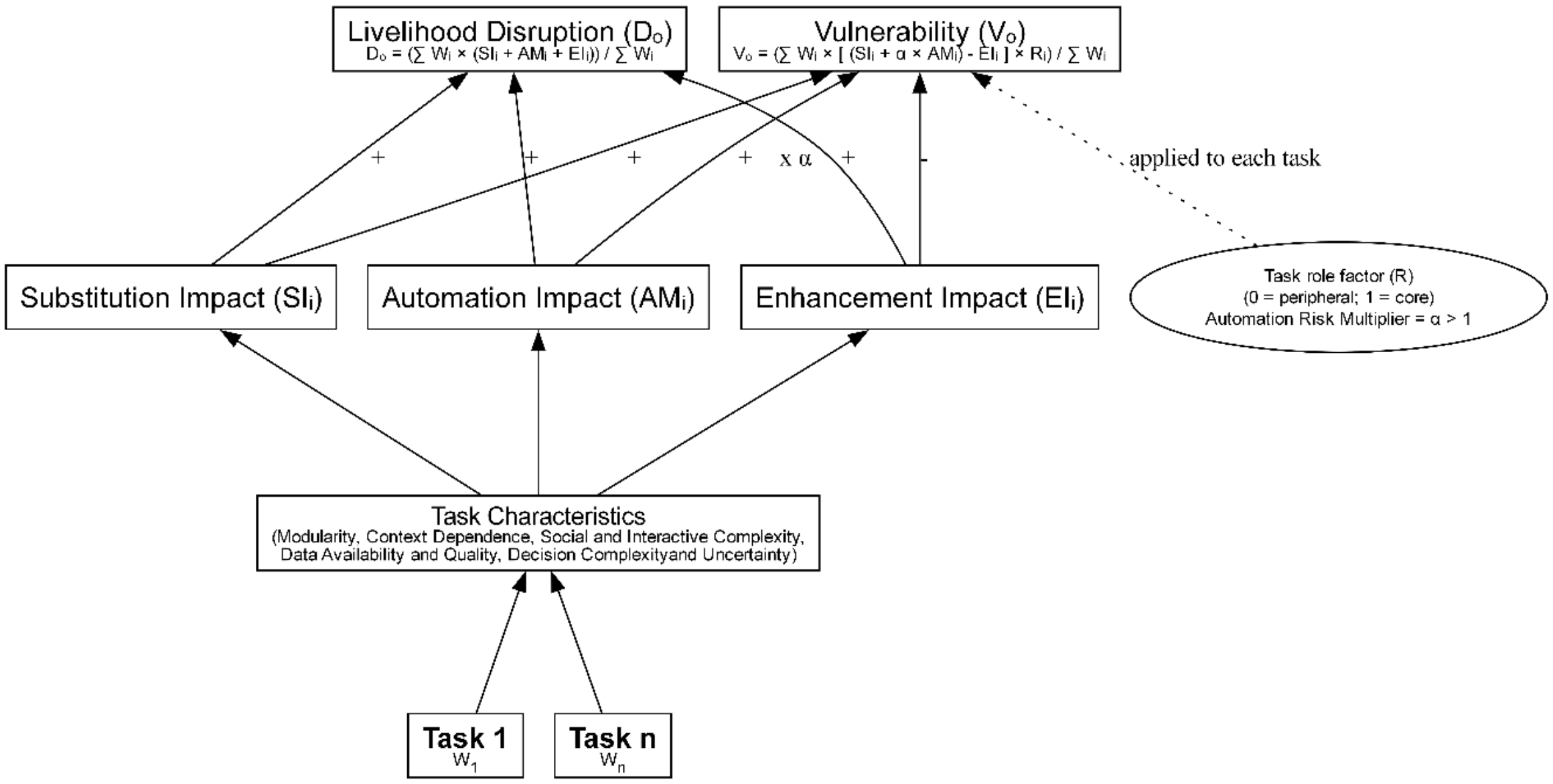

**Figure 1: Task-Based Model to Measure AI-Driven Work Disruption and Livelihood Vulnerability in Freelance Work**

### Task Characteristics Examination

The approach to examining the work task characteristics draws on seminal works (Autor, 2015; Autor et al., 2003; Goodhue & Thompson, 1995) and more recent contributions (Acemoglu & Restrepo, 2019; Frey & Osborne, 2017). Earlier studies have primarily relied on dichotomous or fixed criteria, such as those distinguishing between routine[5] versus non-routine or manual versus cognitive tasks, to gauge the computerization-driven automation potential of occupations (Autor, 2015; Autor et al., 2003).

However, emerging research suggests that the increased capabilities of AI and the blurring of boundaries between such legacy distinctions require a more nuanced approach (Acemoglu & Restrepo, 2019;

[5] Routine tasks adhere to established, clearly defined protocols that enable them to be codified and executed automatically through algorithmic processes (Autor et al., 2003)

Frey & Osborne, 2017). The framework presented above primarily draws on this emerging body of literature. The framework proposes assessing occupational tasks associated with different categories of platform work, including unpaid periphery tasks, along the following five dimensions.

#### *Task Modularity and Granularity*

Prediction tasks that are highly modular and discrete may be easier to automate fully, leading to a strong substitution effect. In contrast, tasks interwoven with other activities might be less amenable to complete automation but could benefit from AI augmentation. This conceptualization builds on the insights from the Task Technology Fit literature (Goodhue & Thompson, 1995) and the well-acknowledged routine versus non-routine and cognitive versus manual task distinctions by Autor and Murnane (Autor, 2015; Autor et al., 2003) to explain how technological change affects job tasks.

#### *Context Dependence*

Tasks that require high sensitivity to environmental, cultural, or situational cues are less likely to be entirely substituted by AI, even if certain components can be automated. In contrast, low context-dependent tasks might face higher substitution as prediction and decision-making abilities can both be performed by AI tools (Acemoglu & Restrepo, 2019; Agrawal et al., 2019).

#### *Social and Interactive Complexity*

Tasks that involve significant social interaction, negotiation, or emotional labour are more likely to experience enhancement, where AI tools support human decision-making, rather than outright replacement (Brynjolfsson et al., 2021).

#### *Data Availability and Quality*

The extent to which a task is data-driven plays a crucial role. Tasks with abundant, high-quality data can be more effectively learned and replicated by AI systems, increasing the likelihood of automation and substitution, whereas tasks reliant on tacit knowledge may be less affected (Agrawal et al., 2019). The lower quality of data may result in unreliable performance and utility of AI tools (Mohammed et al., 2022) resulting in a greater need for human decision-making.

#### *Decision Complexity under Uncertainty*

Finally, tasks that involve multi-step decision-making under uncertainty may be only partially automatable, as AI may offer decision support and augment human performance, rather than fully substitute or automate it. Even as AI systems improve their ability to forecast outcomes, the difficulties of consecutive decision-making under uncertainty require human intervention to manage residual ambiguity (Brynjolfsson et al., 2018; Huang & Rust, 2021).

### The Dimensions of AI Impact

In developing this model, I heavily draw on the dimensions of AI impact as conceptualized by Agrawal et al., (2019). Their framework differentiates between the three direct effects mentioned below, through which advances in AI's prediction capability could influence the future of work. These three dimensions are the building blocks for this task-based impact measurement model. In this framework, each task within an occupation is evaluated along substitution, automation, and enhancement vectors, and their relative importance is adjusted using two key parameters, i.e., task weightage ($W$) and role factor ($R$).

#### *Substitution Impact (SI)*

AI could directly substitute capital for labour in prediction tasks that depend on using existing data to fill information gaps (Agrawal et al., 2019). For instance, freelance transcription platforms increasingly employ AI-powered speech-to-text tools that substitute for human transcribers. As the cost of high-quality prediction decreases, AI tools can replace human labour in the bulk of this work, thereby reducing the need for human labour to do the *entire* work, boosting productivity and shrinking human labour demand. However, I suggest that, for substitution to be viable, enabling factors such as data availability and task modularity must surpass a minimum level. Moreover, excessive uncertainty may encompass high decision complexity and contextual and social challenges that make it difficult for AI to operate reliably, thereby reducing its substitution potential.

#### *Automation Impact (AM)*

When automated prediction increases the relative returns to capital versus labour in complementary decision tasks, it can lead to the complete automation of those tasks (Agrawal et al., 2019). For instance, for micro-tasks, algorithms and AI agents may themselves complete the job or validate task completion, effectively automating decisions that were previously made by human supervisors. In such cases, AI not only predicts outcomes more quickly than a human but also automates the subsequent decision-making process, ultimately reducing the need for the worker. In this context, I suggest that for automation, not only must data availability and the modular nature of tasks be sufficiently high, but they must often exceed a minimal threshold level to enable AI to perform both prediction and decision-making. Excessive levels of hindering factors (e.g., context dependence and social complexity) can drastically reduce the automation potential for AI to take over tasks, regardless of how strong the enabling conditions might otherwise be.

#### *Enhancement Impact (EI)*

Alternatively, the impact of AI on tasks may enhance, rather than replace, human decision-making when it does not replace labour with capital (i.e., AI). When the enabling factors (such as data quality and availability, and the modularity of a task) exist but do not reach the threshold required to achieve substitution or automation levels, it will likely result in AI being a helpful tool for labour without impacting wages.

I suggest that enhancement without substitution tends to be most effective within a balanced range of conditions. When enabling factors are too low, AI adds little value to tasks. On the other hand, when enabling factors are very high, particularly alongside strong hindering factors such as extreme social complexity, tasks may become so unstructured that even augmentation becomes difficult. In this case, by providing accurate, real-time recommendations, AI can augment human productivity in prediction and decision-making tasks without diminishing the value of human labour. For instance, freelance writers can utilize AI-based grammar and style checkers to refine their drafts, thereby enhancing efficiency while maintaining creative control. Here, the prediction provided by AI enables a more nuanced and effective human response, thereby increasing labour productivity.

### Unpaid Work and Task Significance

Not all tasks within an occupation hold the same value[6], particularly in freelance work. To account for the significance of each task, the framework assigns a weight (W) that reflects its significance in contributing to overall work output. Additionally, the role factor (R) is employed to adjust for the economic impact of individual tasks, ranging from unpaid work that is considered peripheral functions (R close to 0) to those that are fully income-generating (R equal to 1).

Traditional occupational analyses often overlook a critical dimension of platform work, the unpaid labour inherent in platform-mediated environments (Bonina et al., 2021; Cingolani, 2021; Standing, 2017). Unpaid work tasks, such as developing proposals and applying for jobs, a constant need for upskilling to complete new work, and a range of project management responsibilities that come with being a freelance worker, are essential for securing and executing paid work but are rarely captured in standard job descriptions. The role of unpaid labour in the job quality of Kenyan platform workers is also important, since workers in developing regions do a disproportionate amount of unpaid labour (Howson et al., 2023). Recognizing the significance of this gap, this model incorporates these hidden tasks alongside conventional work tasks to provide a more comprehensive picture of occupational dynamics and the impact of AI on these tasks.

### AI-Induced Livelihood Disruption

The disruption construct in this framework encompasses any change from the status quo, even if such change is perceived as beneficial, as it necessitates new skills (e.g., learning skills to utilize AI) or adaptation. In this model, disruption is driven by the combined effects of substitution *(SI)* and automation *(AM)* impacts, as well as by any worker enhancement *(EI)* in the sense that even changes beneficial to a worker require adjustment. Disruption score ($d_i$) for each task within an occupation is calculated as:

$$d_i = (1 - SI_i) \times (1 - AM_i) \times (1 - EI_i)$$

Measuring the disruption score using a multiplicative (complementary) method captures the combined impact of its enabling factors, ensuring that the overall disruption score will be high if any single enabling factor is high. The disruption score ($D_o$) at the occupation level is calculated by aggregating the weighted task disruption scores as follows:

$$D_o = \frac{\sum_{i=1}^{n} W_i d_i}{\sum_{i=1}^{n} W_i} = \frac{\sum_{i=1}^{n} W_i \times [(1 - SI_i) \times (1 - AM_i) \times (1 - EI_i)]}{\sum_{i=1}^{n} W_i}$$

Note: Here, di is the disruption score for a task. Wi is the weight assigned to the task. SIi, AMi, and EIi are the Substitution, Automation, and Enhancement impact scores for task i, respectively. n is the total number of tasks within the occupation.

### AI-Induced Livelihood Vulnerability

Vulnerability captures the net negative impact on an occupation or work category. Specifically, the risk that the expansion of AI may influence work availability and remuneration for completing that work. It focuses on the adverse effects of substitution and automation on workers while recognizing that enhancement may, to some degree, mitigate these adverse outcomes by complementing human labour. Thus, for each task, the vulnerability contribution is derived by subtracting the mitigating effect of enhancement from the combined disruptive effects. Rather than viewing these scores as absolute measures of a task's automation, substitution, or enhancement potential, I interpret the final vulnerability score as a probabilistic indicator of risk. The vulnerability score ($v_i$) for each task within an occupation is calculated as:

$$v_i = (Weighted\,SI_i + \alpha \times Weighted\ AM_i) \times \frac{R_i - 1}{R_{max} - 1} - Weighted\ EI_i - \times \beta \times (1 - \frac{R_i - 1}{R_{max} - 1})$$

[6] In this context, value refers to the relative importance or contribution of a task to the overall work output within an occupation. This value may be economic (e.g., income-generating potential or client willingness to pay), effort-based (e.g., time intensity or complexity), or skill-based (e.g., level of expertise or specialized knowledge required). It recognizes that not all tasks are equal in their demands or returns, particularly in freelance settings where workers often engage in a mix of high-skill and routine tasks.

Subsequently, the overall vulnerability score for an occupation ($V_o$) is computed as:

$$V_o = \frac{\sum_{i=1}^{n} W_i v_i}{\sum_{i=1}^{n} W_i} = \frac{\sum_{i=1}^{n} W_i \times (Weighted\,SI_i + \alpha \times Weighted\ AM_i) \times \frac{R_i - 1}{R_{max} - 1} - Weighted\ EI_i - \times \beta \times (1 - \frac{R_i - 1}{R_{max} - 1})}{\sum_{i=1}^{n} W_i}$$

Here $\alpha$ is a multiplier, $\alpha>1$ may increase the weight of the Automation Impact score, based on the notion that automation tends to have a more significant adverse effect on job security than substitution[7]. This research proposed setting the initial $\alpha$ value at 1.5. $R_i$ is the role factor for task $i$. This factor adjusts the negative impact of a task based on its monetary significance (unpaid tasks offset automation and substitution impact). β scales the impact of the portion of the task less directly tied to income (to account for the positive impact of unpaid work in this case) on the vulnerability calculation.

### Limitations

While this framework offers a detailed process for dissecting and analyzing AI's impact on various platform work categories, it has several limitations. First, measuring latent variables, such as disruption and vulnerability, inevitably involves subjective judgment, which can introduce variability in interpreting results. Second, accurately determining task weightings and assessing task characteristics requires a systematic and context-specific approach, as the nuances of each occupation must be carefully examined. Nevertheless, the granularity of this framework is also a strength, as it enables the integration of rich contextual information into the model, thereby enhancing its overall relevance and precision.

## *Methods*

This study employs a mixed-methods design to operationalize the Task-Based Model for Measuring AI-Driven Disruption and Vulnerability in the Platform Work Economy, introduced earlier. The identification of paid and unpaid tasks across the selected occupational categories was iteratively completed by reviewing online job listings on Upwork, Freelancer, and participant portfolios (n = 22), as well as relevant literature on identifying unpaid tasks between March 2024 and July 2024. These categories included freelance writing, virtual assistance, web and software development, and transcription services.

In the quantitative modelling component, identified tasks for these occupations were first analyzed and scored against their relative significance (weight) and task characteristics criteria. The Analytic Hierarchy Process (AHP) was employed to quantify task characteristics and assign weightage scores, resulting in impact scores across the three dimensions of AI exposure (i.e., automation, substitution, and enhancement). AHP, as a multi-criteria decision-making tool, is well-suited for scenarios like this research, where subjective judgments about weightage must be rationally quantified. AHP breaks down the decision-making process into a hierarchy of more easily comprehended sub-problems that can be analyzed independently (Brunelli, 2014; Salomon & Gomes, 2024; Vargas, 1990). This method is particularly suitable for this study, as it provides a clear and logical framework for comparing the relative importance of different criteria to determine weightages. Utilizing the processes presented in the task-based framework, this analysis generated two key metrics: a disruption score, which quantifies the magnitude of AI impact, and a vulnerability score, capturing the qualitative aspects of AI impact on primary income-generating and unpaid auxiliary tasks for each occupation.

### Task Identification, Weightage, and AI Impact Calculation

In the first step, core and supporting tasks within each occupation were identified by reviewing job postings on freelance platforms and examining

[7] Automation often implies the removal of the human element entirely and could replace entire processes rather than merely altering tasks. This is supported in literature such as Acemoglu and Restrepo (2018), who highlight that automation can lead to significant labour displacement. A multiplier greater than 1 (i.e., $\alpha>1$) is a way to capture that asymmetry but could be adjusted based on empirical observations and sensitivity analysis

participant portfolios, ensuring the selected tasks accurately reflect available opportunities and actual freelancer experiences. Note that the occupations included are not exhaustive of the platform work economy.

Leveraging the conceptual framework and criteria presented earlier, task characteristics (i.e., task modularity and granularity, context dependence, social and interactive complexity, data availability and quality, and decision complexity under uncertainty) were scored on a scale of 1–10 through a series of calls to context-informed GPT-4 model between September 2024 and January 2025. This scoring process was also used to compute the task weightage role factor (R), which accounts for the contribution of unpaid tasks. These initial scores were then manually verified and calibrated based on a review of empirical data and observations. This included adjusting context information where model outputs diverged from grounded realities, such as revising the unpaid task ratio based on direct feedback from platform workers and cross-checking the listed tasks against documented work samples and field interviews.

This method follows the previous research by Gmyrek et al. (2023). Gmyrek et al. (2023) utilized ChatGPT4 API prompts to generate work task scores, predicting the automation potential of generative AI. In their work, the model was prompted to analyze a uniform set of tasks for each occupation, after which the outputs were used to calculate mean automation scores. In contrast, my methodology employs a similar context-specific prompt strategy. However, rather than relying on AI judgment to gauge a task's automation potential, it focuses on scoring task characteristics and pairwise matrix relationships against a refined set of theoretically grounded criteria to mitigate technology capability biases and enhance replicability and transparency (Karger et al., 2023). For detailed task scoring against task characteristics with clear justification, see Appendix D. For detailed weightage and paid-unpaid task ratio scoring, see Appendix E.

While some elements of scoring involved subjective judgment, these judgments are grounded in empirical observations I performed during my fieldwork in Kenya, direct insights from Upwork and Freelancer, where I registered and applied for jobs as a platform worker to gain a better insight into validated job postings, and interviews with workers (n=20). Besides the empirical evidence-driven adjustments, a sensitivity analysis was conducted to ensure the robustness of the model.

## AI Impact Calculation using AHP

A hierarchical model using the Analytic Hierarchy Process (AHP) was developed to assess the potential impact of AI on tasks within an occupation along three dimensions: Substitution Impact (SI), Automation Impact (AM), and Enhancement Impact (EI). These dimensions quantify how AI will likely affect tasks by measuring both the overall disruption and the net vulnerability. The following steps were taken to calculate different dimensions of AI impact, i.e., automation (AM), substitution (SI), and enhancement impact (EI) score for each task.

### *Step 1: Hierarchy*

As a first step, a hierarchy was developed to structure the decision-making process into levels:

**Level 1 (Goal):** Determine the automation, substitution, and enhancement impact of AI on each task.

**Level 2 (Criteria):** Identify and compare the task characteristics that influence AI impact dimensions *(e.g., SI, AM, EI)*. The five key characteristics are:

*Task Modularity and Granularity (TMG)*: Highly modular and discrete tasks can be automated or substituted by AI. In contrast, tasks that are interwoven with other activities may require higher human judgment and thus be less amenable to full automation.

*Context Dependence (COD):* Tasks that rely on sensitivity to environmental, cultural, or situational cues are less susceptible to full automation as they require human judgment, whereas those with low context dependence can be more easily replaced by AI.

*Social and Interactive Complexity (SIC):* Tasks requiring significant social interaction, negotiation, or emotional labour tend to be either enhanced through AI support or are not impacted by AI.

*Data Availability and Quality (DAQ):* Tasks supported by abundant, high-quality data are more amenable to automation, while those dependent on tacit knowledge or lower-quality data may continue to require human judgment.

*Decision Complexity under Uncertainty (DCU):* Tasks involving multi-step decision-making under uncertainty are likely only partially automatable but may still see significant substitution. AI can provide decision support, but human intervention remains essential to address residual ambiguities. Tasks with simple, rule-based decisions are easier to automate.

**Level 3 (Alternatives):** The individual core tasks within the occupation.

***Step 2: Pairwise Comparison of Criteria***

For each AI Impact dimension (SI, AM and EI), a pairwise comparison of the five criteria (i.e., task modularity and granularity, context dependence, social and interactive complexity, data availability and quality, and complexity under uncertainty) is conducted to determine their relative importance. Each pairwise matrix is normalized by dividing each element by its respective column total.

$$Normalized\ Value_{ij} = \frac{Element\ Value_{ij}}{Column\ Total_j}.$$

The relative weights $w_i$ of each criterion i are calculated by taking the average of the normalized values in its row, which yields a set of relative weights that reflect the judgments in the pairwise comparisons.

$$w_i = \frac{\sum_{j=1}^{n} Normalized\ Value_{ij}}{n}$$

Here n is the number of criteria

***Step 3: Consistency Check***

A consistency check is performed to ensure the reliability of the judgments. The weighted sum vector (*s*) is calculated by multiplying each row of the original pairwise comparison matrix by the corresponding criterion weight (*w)*. This yields the weighted sum vector, where each element represents the aggregated importance for a criterion.

$$S_i = \sum_{i=1}^{n} a_{ij} \times w_j, for\ i = 1,2, \dots, n.$$

For each criterion, the consistency vector element ($\lambda_i$) is calculated by dividing the weighted sum for that criterion by its corresponding weight.

$$\lambda_i = \frac{s_i}{w_i}$$

The Consistency Index (CI) is calculated as: $CI = \frac{\lambda max - n}{n-1}$

Where λmax is the largest eigenvalue of the pairwise comparison matrix calculated as

$$\lambda_{max} = \frac{1}{n}\sum_{i=1}^{n} \lambda_i$$

n is the number of criteria.

The Consistency Ratio (CR) is calculated as $CR = \frac{CI}{RI}$

RI is the random consistency index, which is determined based on the number of criteria. If CR < 0.10, the judgments in the pairwise comparison matrix are considered consistent.

***Step 4: Incorporating Threshold Effects using Gaussian Functions***

I incorporate threshold adjustments via Gaussian functions to better reflect the nuanced conditions under which AI impacts tasks. It reflects the notion that each criterion has an optimal level, and any deviation (either too low or too high) results in a proportionally lower contribution. Before calculating the weighted impact scores for each AI impact dimension, the raw scores (on a 1–10 scale) for each criterion are transformed into effective scores using a Gaussian transformation.

$$f_i(r_i) = \exp(-\frac{(r - r_{opt})^2}{2\sigma^2}$$

Here, $r_i$ is the raw score of a particular criterion. σ is the standard deviation that controls the width of the sweet spot.

This transformation peaks at 1 when the raw score equals the optimal value ($r_{\text{opt}}$) and decays exponentially as the raw score deviates from this optimum. This Gaussian transformation (with a sigma value of 1.5, determined via sensitivity analysis results below) ensures that tasks far from the ideal conditions receive substantially lower effective scores. This adjustment captures the concept of the sweet spot, as explained in the conceptual framework section, reflecting that only scores near the optimum contribute maximally to the subsequent impact calculations.

For each AI impact dimension, optimal values are set based on theoretical expectations and empirical insights into task structure and complexity. Substitution is most effective when tasks are highly modular and supported by robust data while exhibiting minimal contextual, social, and decision complexity. Therefore, for factors that enable automation and substitution potential, such as Task Modularity and Data Availability, I set $r_{opt}$ = 8 (on a scale of 1 to 10). This reflects the expectation that tasks scoring near 8 on these dimensions are ideally structured for AI to fully substitute human labour. Conversely, lower values facilitate substitution for hindering factors (Context Dependence, Social and Interactive Complexity, and Decision Complexity). I set $r_{opt}$ = 2 for these criteria, meaning that tasks ideally should score close to 2 to minimize the need for human judgment. Deviations from these optima result in proportionally lower effective scores, reducing the overall substitution impact.

Automation requires nearly perfect enabling conditions, as the process involves prediction and autonomous decision-making. To reflect this, I set $r_{opt}$ =10 for enabling factors, indicating that only tasks with high modularity and data quality are optimal for full automation. For hindering factors, where even minor increases in complexity can thwart automation, I set $r_{opt}$ =1. These choices underscore that even slight deviations from the ideals, such as a modest rise in contextual or decision complexity, can significantly diminish automation potential.

Enhancement occurs when AI augments rather than replaces human labour (Acemoglu & Restrepo, 2019). This dimension is most effective when tasks possess a balanced mix of structure and complexity; that is, when enabling factors are present but not at levels high enough to favour substitution or automation, and hindering factors are moderate enough to require human judgment without overwhelming the AI's supportive role. Accordingly, I set $r_{opt}$ =5 for both enabling and hindering factors in the context of enhancement. A score of 5 represents a mid-range condition where AI can effectively augment human performance.

#### *Step 5: Weighted Task Score Calculation for AI Impact Dimension*

Effective scores are combined with the AHP-derived weights to compute the impact score for each task. This process is repeated separately for *SI, AM, and EI.* The relative weight from the pairwise matrix is then applied to calculate the weighted scores. For each AI impact dimension $X$ (where $X$stands for SI, AM, or EI), the weighted impact score for a task is computed as:

$$\text{Weighted Imapct Score}_{\text{X}} = \sum_{i=1}^{5} (f_i(r_{i,X}) \times w_i)$$

Where X represents the impact dimension (SI, AM, or EI), $r_{i,X}$ is the score assigned for the $i_{\text{th}}$ criterion for dimension $X$, and $w_i$ is the corresponding weight. This process results in three separate weighted scores per task: weighted substitution impact (SI), weighted automation impact (AM), and weighted enhancement impact (EI).

#### *Disruption and Vulnerability Scores*

Using the computed weighted impact scores for SI, AM, and EI, I derived the overall task-level impact measures. Disruption in the model is calculated as a multiplicative (complementary) method and represents the combined impact of three abstract dimensions, i.e., substitution, automation, and enhancement.

$$d_{task} = 1 - [(1 - Weighted\,SI)(1 - Weighted\,AM)(1 - Weighted\,EI)]$$

This multiplicative method ensures that the combined disruption score is sensitive to the highest individual impact and reflects the interdependent, competing nature of the risks posed by substitution, automation, and enhancement.

Vulnerability captures the net negative (economic) impact on the occupation. Since automation tends to have a more detrimental effect, an automation multiplier α (α=1.5) is applied to the automation scores to compute vulnerability. In addition, to reflect that not all tasks are equally economically important, a role factor $R$ (derived from the task identification process) is incorporated, where $R$ is closer to 1 for core or income-generating tasks and less than 1 for supportive tasks. The vulnerability score for a task is calculated as:

$$v_{task} = (SI + \alpha\ \times AM) \times \frac{R-1}{R_{max}-1} - Weighted\,EI\ - \times \beta\ \times (1 - \frac{R-1}{R_{max}-1})$$

Raw scores for vulnerability can range from a potentially protective extreme (negative) to a highly vulnerable extreme (positive). In the results, this score is normalized to a [−1, +1] scale for clarity. Specifically, −1 indicates the theoretical maximum advantage (where enhancement strongly outweighs substitution and automation), while +1 reflects the theoretical maximum vulnerability (where substitution and automation dominate).

### Task weightage using AHP

A key distinction I make in this research is that each task within a given occupation may hold a different weightage in terms of its significance and the value it adds to the overall workload of a platform worker. In this research, I theorize that this significance and value are dependent on four key factors i.e., frequency (how often a task is performed in a particular category of platform work), duration (the time required for a task), complexity (the skill and effort needed), and client prioritization (the task's importance from the client's perspective).

Pairwise comparisons of these factors were conducted to assess their relative importance, yielding a set of weighted values. Each core task was then scored against these criteria, and the scores were multiplied by the corresponding weights. Finally, the aggregated scores were normalized to produce percentage weightings that reflect the relative significance of each task within the occupation. This structured approach ensures a transparent, reproducible, and justifiable means of prioritizing tasks in line with the role's demands.

#### *Task Weightage AHP Step 1: Hierarchy*

As a first step, I developed a hierarchy that structures the decision-making process into levels: the overall goal, the criteria, and the alternatives (i.e., the core tasks within each occupation).

**Level 1 (Goal):** Determine the weightage for each core task.

**Level 2 (Criteria):** Frequency (F), Duration (D), Complexity (C), Client Prioritization (P). Where Frequency (F) reflects how often a task is performed in the freelancer's workflow; Duration (D) reflects the average amount of time required to complete a task; Complexity (C) reflects the perceived cognitive and technical difficulty of the task; and Client Prioritization (P) reflects how important the task is seen by clients in the platform economy (from a workers' perspective.

**Level 3 (Alternatives):** The core task identified for each occupation.

***Step 2: Pairwise Comparison of Criteria***

Next, a pairwise comparison of the criteria (*F, D, C, P*) is conducted. The aim is to create a pairwise matrix to compare the criteria based on their relative significance in the context of the specific occupation while ensuring that the relative importance of the criteria is grounded in the realities of the platform economy (as compared to traditional job descriptions).

To derive the relative weights of the criteria, the pairwise matrix is normalized by dividing each element of the matrix by its respective column total, thereby normalizing the matrix.

$$Normalized\ Value_{ij} = \frac{Element\ Value_{ij}}{Column\ Total_j}.$$

Finally, the relative weight of each criterion *I* is calculated by taking the average of the normalized values in its row, yielding a set of relative weights that reflect the expert judgments expressed in the pairwise comparisons.

$$Weight\ of\ Criterion_i = \frac{\sum_{j=1}^{n} Normalized\ Value_{ij}}{n}$$

Where n is the number of criteria.

Step 3: Consistency Check

A consistency check is performed to ensure the reliability of the judgments made. In this step, the Weighted sum vector (*s*) is calculated by multiplying each row of the original pairwise comparison matrix by the corresponding criterion weight (*w*). This yielded the weighted sum vector, where each element represented the aggregated importance for a criterion.

$$s_i = \sum_{i=1}^{n} a_{ij} \times w_j, for\ i = 1,2,\ldots,n.$$

For each criterion, I calculated the consistency vector element ($\lambda_i$) by dividing the weighted sum for that criterion by its corresponding weight.

$$\lambda_i = \frac{s_i}{w_i}$$

The Consistency Index (CI) is calculated as:

$$CI = \frac{\lambda max - n}{n - 1}$$

λmax is the largest eigenvalue of the pairwise comparison matrix calculated as:

$$\lambda_{max} = \frac{1}{n}\sum_{i=1}^{n} \lambda_i$$

n is the number of criteria.

The Consistency Ratio (CR) is calculated as $CR = \frac{CI}{RI}$

Here, RI is the random consistency index, which is determined based on the number of criteria. If CR < 0.10, the judgments in the pairwise comparison matrix are considered consistent.

***Step 4: Weighted Task Score Calculation***

The pairwise table is then used to calculate the weighted scores for each task within each occupation. This step quantifies the overall significance of each task based on its performance across the criteria. These scores are subsequently normalized and converted into percentage values. The weighted score for each task is calculated using the following formula:

$$\text{Weighted Score for Task} = (\text{F score} \times \text{F weight}) + (\text{D score} \times \text{D weight}) + (\text{C score} \times \text{C weight}) + (\text{P score} \times \text{P weight})$$

The weighted scores are normalized by dividing each task's weighted score by the total score for all tasks:

$$\text{Normalized Score for Task} = \frac{\text{Weighted Score for Task}}{\text{Total Weighted Score for All Tasks}} \times 100$$

The collective results of this process ensured that the assigned weightages for tasks within each occupation are systematic, justified, and consistent with the underlying criteria.

## Occupation Level Impact Scores

Scores across all tasks within the occupation were aggregated using this task weightage to capture occupation-level impact. The occupation-level measures are expected to shed light on the composite constructs of disruption and vulnerability, recognizing the individual dimensions of AI impact (SI, AM, and EI) and differences between occupations. Such disaggregated information is helpful for understanding which aspect of AI impact (e.g., substitution vs. automation) is most pronounced within an occupation. Scores are aggregated at the occupation level and normalized to a scale of 0 to 1, providing insight into the relative contributions of each AI impact dimension across occupations.

$$SI_o = \frac{\sum_{i=1}^{n} W_i \times SI_i}{\sum_{i=1}^{n} W_i} \quad AM_o = \frac{\sum_{i=1}^{n} W_i \times AM_i}{\sum_{i=1}^{n} W_i} \quad EI_o = \frac{\sum_{i=1}^{n} W_i \times EI_i}{\sum_{i=1}^{n} W_i}$$

SIo is Occupation-Level Substitution Impact, and SIi is the weighted Substitution Impact score for each task i
AMo is the Occupation-Level Substitution Impact, and AMi is the weighted Substitution Impact score for each task i. EIo is Occupation-Level Substitution Impact and AMi is the weighted Substitution Impact score for each task i. Wi the task weightage (obtained from the Task Weightage AHP process).

Using the task-level disruption and vulnerability scores, occupation-level measures for Disruption and Vulnerability are computed as:

$$D_o = \frac{\sum_{i=1}^{n} W_i d_i}{\sum_{i=1}^{n} W_i} = \frac{\sum_{i=1}^{n} W_i \times [(1 - SI_i) \times (1 - AM_i) \times (1 - EI_i)]}{\sum_{i=1}^{n} W_i}$$

$$V_o = \frac{\sum_{i=1}^{n} W_i v_i}{\sum_{i=1}^{n} W_i} = \frac{\sum_{i=1}^{n} W_i \times (Weighted\ SI_i + \alpha \times Weighted\ AM_i) \times \frac{R_i - 1}{R_{max} - 1} - Weighted\ EI_i - \times \beta \times (1 - \frac{R_i - 1}{R_{max} - 1})}{\sum_{i=1}^{n} W_i}$$

## ***Consistency and Robustness Checks***

All AHP pairwise comparisons yielded CR values below 0.10, indicating acceptable consistency (Salomon & Gomes, 2024). Specifically, for Freelance Writing, the CRs were 0.072 for the substitution impact, 0.039 for the automation impact, 0.067 for the enhancement impact, and 0.026 for the task weightage. For Freelance Virtual Assistance, the CRs were 0.072 (substitution), 0.039 (automation), 0.067 (enhancement), and 0.017 for the task weightage. In Freelance Transcription, the CRs were 0.071, 0.040, 0.067, and 0.026, respectively, while for Freelance Developers they were 0.072, 0.040, 0.067, and 0.043. These results confirm that the input judgments in the AHP comparisons are internally consistent.

To assess the robustness of the vulnerability metric with respect to the sigma parameter in the Gaussian transformation, I conducted a sensitivity analysis across four occupational groups. The analysis shows that as σ increases from 0.5 to 3.0, the overall paid work vulnerability score rises dynamically. Notably, at σ = 1.5, the vulnerability scores are significantly lower, reflecting a steeper decay in effective task scores when raw scores deviate from their optimum. This steep decay supports the theoretical expectation that even minor deviations markedly reduce automation potential. Consequently, selecting σ = 1.5 establishes a robust threshold that effectively distinguishes tasks likely to be automated from those that are not.

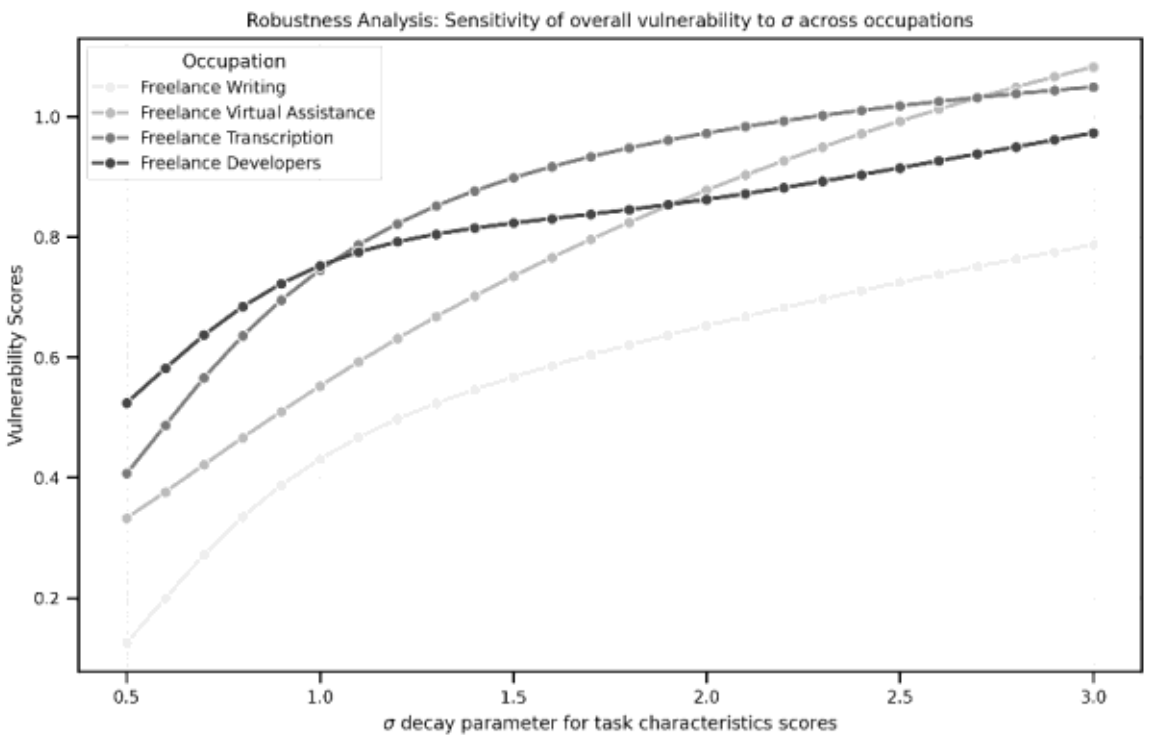

**Figure 2: Sensitivity Analysis of Vulnerability Index to Sigma Parameter**

As shown in Figure 5.2, the sensitivity analysis indicates that as σ, the decay parameter controlling how rapidly the effective score decreases as raw scores for task characteristics deviate from the optimum, increases from 0.5 to 3.0. Overall, these results confirm that while the absolute vulnerability values are sensitive to sigma, as expected given its role in modulating the decay rate, the qualitative trends remain consistent across occupational groups. The modest range of variation, particularly for Freelance Developers, lends confidence to the stability of this model. Consequently, the chosen baseline sigma value (e.g., 1.5) is well justified, and the consistent behaviour across occupations supports the robustness of our analytical framework.

### **Empirical Fieldwork**

The fieldwork was conducted to understand workers' perspectives on the impact of AI on their livelihoods and to validate and enrich the theoretical assumptions underlying the AI impact modelling. The field research data were collected in Kenya from January 2024 to July 2024, using a survey of 52 freelance platform workers, from which 20 participants were selected for follow-up, semi-structured interviews. Survey participants were recruited using purposive and snowball sampling techniques, targeting active freelance platform workers across different sectors to capture a range of experiences and work arrangements in Kenya. The aim was to capture a diverse sample in terms of occupation, gender, experience level, and engagement with AI tools. The survey captured information, including the frequency of generative AI tool use, perceived helpfulness of AI tool use, and perceived job loss concern. For a detailed list of survey questions, please see Appendix A.

Please note that the results of the field study are incorporated into the discussion section, while the results section in this article primarily focuses on the theoretical modelling and projections of AI's impact on selected freelance occupations. This structure was chosen to support clearer interpretation, allowing the theoretical modelling in the results section to be complemented by the field study insights in the discussion section.

#### *Fieldwork Data Analysis*

Descriptive statistics were used to profile and cross-tabulate key indicators by occupation. Bivariate analyses, including t-tests and Pearson correlations, explored associations among AI use, job-loss concern, and labour input variables. An exploratory factor analysis using principal-axis factoring with oblique rotation was conducted to assess whether AI-use frequency and perceived helpfulness could be meaningfully combined into a single AI-Utility index, based on acceptable factor loadings and internal reliability. Ordinary least squares (OLS) regressions were then conducted to examine two outcomes: first, whether AI-Utility predicts job-loss concern; and second, which factors, including AI concern, paid hours, age, gender, education, platform tenure, work control, and financial satisfaction, explain workers' confidence in securing future platform work. All regression models were tested for linearity, homoscedasticity, and multicollinearity (VIF < 5), and confirmed to have normally distributed residuals. Bootstrap standard errors, based on 1,000 replications, were used to improve the robustness of coefficient estimates. This analytical sequence provided a rigorous quantitative foundation for interpreting how Kenyan freelancers perceive and respond to AI-driven changes in the platform labour market.

A narrative approach was used to examine and contextualize workers' accounts of how AI is reshaping their work. Interview recordings were transcribed verbatim and reviewed for accuracy and clarity. A narrative approach is a qualitative method that examines how individuals construct meaning through the stories they tell, focusing on both the content and structure of their accounts within broader social and cultural contexts (Riessman, 2008). The analysis focused on how participants constructed meaning around AI through personal experiences, language choices, and the sequencing of events. This approach made it possible to capture not only what workers said about AI, but also how they positioned themselves within evolving technological landscapes and shifting labour dynamics. Direct quotes from the interviews are used, and all participant names are pseudonyms to protect their identities.

## *Results*

The analysis reveals that AI-driven work reconfiguration has a significant impact on all platform work categories (Table 5.1). The overall disruption factor registers an exceptionally high average score of 0.913, with paid (0.906) and unpaid work (0.929) exhibiting similarly elevated means, indicating the substantial ongoing and anticipated transformation in platform work practices. Various dimensions of this impact, automation (mean = 0.534), enhancement (mean = 0.542), and substitution (mean = 0.649, all yield high values, reflecting the widespread and dual-faceted (both positive and negative) influence of AI. Notably, vulnerability emerges as a critical factor: while the overall vulnerability measure remains relatively low (mean = 0.214), unpaid work vulnerability presents a more pronounced negative value (mean = –0.493), highlighting distinct dynamics in unpaid labour contexts.

**Table 1: Impact of AI on Freelance Platform Work (Multiple Occupation Categories)**

| Variable | Mean | Minimum | Maximum | SD |
|---|---|---|---|---|
| Vulnerability | 0.214 | 0.054 | 0.368 | 0.129 |
| Paid Work Vulnerability | 0.482 | 0.324 | 0.662 | 0.145 |
| Unpaid Work Vulnerability | -0.493 | -0.657 | -0.411 | 0.111 |
| Disruption | 0.913 | 0.887 | 0.947 | 0.025 |
| Paid Work Disruption | 0.906 | 0.891 | 0.944 | 0.025 |
| Unpaid Work Disruption | 0.929 | 0.878 | 0.96 | 0.038 |
| Automation | 0.534 | 0.411 | 0.682 | 0.132 |
| Unpaid Work Automation | 0.464 | 0.334 | 0.603 | 0.145 |
| Enhancement | 0.542 | 0.478 | 0.644 | 0.078 |
| Paid Work Enhancement | 0.496 | 0.408 | 0.627 | 0.101 |
| Unpaid Work Enhancement | 0.654 | 0.627 | 0.714 | 0.041 |
| Substitution | 0.649 | 0.601 | 0.675 | 0.033 |
| Paid Work Substitution | 0.646 | 0.616 | 0.696 | 0.035 |
| Unpaid Work Substitution | 0.648 | 0.573 | 0.724 | 0.077 |

### Occupation Level Impact Results

At the occupational level, automation risk is notably higher for jobs characterized by routine and structured execution, such as web and software development (0.68) and transcription (0.61), reflecting their greater susceptibility to automation (Figure 5.3). In contrast, types of work that require complex decision-making and adaptability, such as freelance writing (0.41) and virtual assistance (0.44), exhibit a considerable but comparatively lower risk, indicating greater resilience against automation.

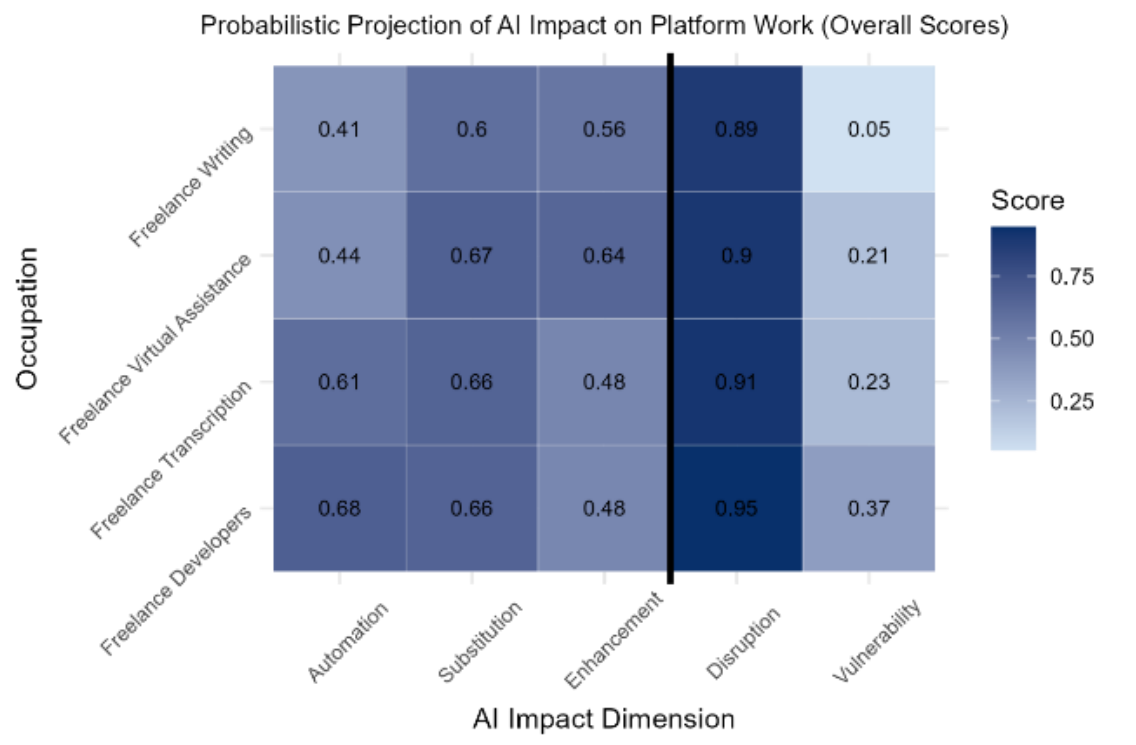

**Figure 3: AI Impact on Platform Work (Overall Scores)**

As highlighted in Figure 5.3, at the occupational level, automation risk is particularly pronounced in roles characterized by routine and structured execution, such as web and software development (0.68) and transcription (0.61). This reflects their greater susceptibility to AI-driven automation, as these tasks often involve repetitive processes that machine learning models can efficiently handle. In contrast, occupations requiring greater adaptability and complex decision-making, such as freelance writing (0.41) and virtual assistance (0.44), exhibit lower but notable automation risks. This suggests that while AI may enhance productivity in these roles, complete automation remains less feasible due to the need for creativity, contextual judgment, and human interaction.

Disruption scores across occupations indicate consistently high levels of AI-driven impact, with all four freelance occupations registering values above 0.88. Web and software development (0.95) and transcription (0.91) exhibit the highest disruption scores, underscoring the substantial transformation potential of these fields. Virtual assistance (0.90) and freelance writing (0.89) follow closely, suggesting that while these roles may be relatively less susceptible to full automation, they are nonetheless experiencing profound shifts in task execution, workflow dynamics, and required skill sets. This highlights the pervasive influence of AI across diverse freelance occupations, affecting not only technical and repetitive tasks but also roles traditionally reliant on adaptability and human judgment.

A key insight from this analysis is that AI-driven changes do not uniformly translate into heightened livelihood vulnerability across all occupations. Notably, freelance writing registers a low vulnerability score (0.05), suggesting that AI's overall impact on job security in this field may be relatively marginal. In contrast, transcription (0.23) and virtual assistance (0.21) exhibit higher vulnerability scores, indicating that these occupations face more significant risks due to AI-driven automation. This suggests that while these roles may persist, the increasing integration of AI could significantly alter their demand, potentially reducing human labour needs or reshaping job functions.

Freelance developers exhibit the highest vulnerability score (0.37), indicating that platform workers in this field may face substantial challenges as AI-driven automation reshapes their work. However, the factors that mitigate or amplify this vulnerability warrant closer examination. While substitution and automation risks are significantly higher in these occupations, they do not always correspond to equally high vulnerability. This suggests that other dynamics, such as the demand for specialized skills, adaptability, or the evolution of AI-human collaboration, may influence the extent to which workers experience job displacement or transformation.

**Unpaid Work as a Balancing Factor**

All four occupations examined carry a significant share of unpaid work, although to varying degrees (Figure 5.4). This assessment was based on rating each task within the four occupations for its contribution to paid or unpaid work (R factor), allowing for an occupation-specific calculation of the share of unpaid tasks and the identification of which tasks fall into each category. The proportion of unpaid work varies significantly across occupations, influencing the overall impact of AI. Virtual assistance has the lowest unpaid work share (19.8%), followed by developers (26.2%). In contrast, writing (35.1%) and transcription (31.7%) involve a substantial unpaid workload. These differences highlight how unpaid tasks alter the overall composition of freelance work (see the task-level analyses for more detailed information on specific paid and unpaid tasks). The prominence of these tasks creates a complex dynamic for how AI tools might reshape practices and opportunities in platform-based work.

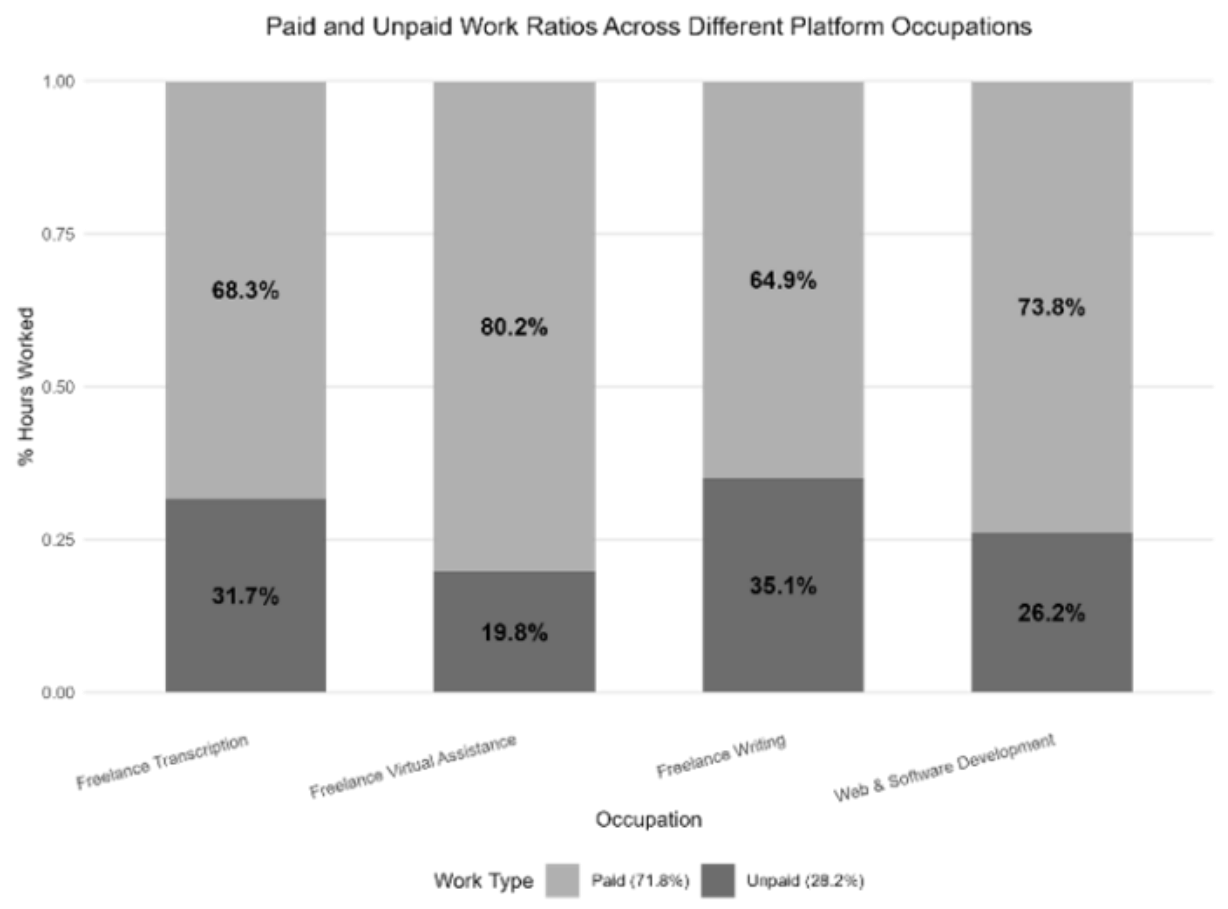

**Figure 4: Paid and Unpaid Work Ratio in Platform Work**

While both paid and unpaid work experience are significantly disrupted by AI, unpaid tasks exhibit a pronounced counterbalancing effect. The results indicate that automation and substitution effects are moderately high across both work types (Automation: 0.56 vs. 0.46; substitution: 0.65). However, enhancement scores are notably higher in unpaid tasks (0.65) than in paid work (0.50). This suggests that AI integration may offer more significant efficiency gains in non-remunerated activities, such as proposal development and client communication.

A key distinction emerges in disruption and vulnerability. Disruption scores remain similarly high for both paid (0.91) and unpaid work (0.93), underscoring the broad impact of AI across task categories. However, vulnerability trends diverge because unpaid tasks counter vulnerability (Figure 5.5). Paid work exhibits a positive vulnerability score (0.48), whereas unpaid work registers a negative score (-0.49), indicating a net offset effect. Despite this mitigating factor, the lower proportion of unpaid work (28.2%) limits its ability to substantially reduce overall vulnerability. These findings suggest that while unpaid tasks may help buffer against AI-driven displacement, their influence is constrained by their relative share within the overall work structure.

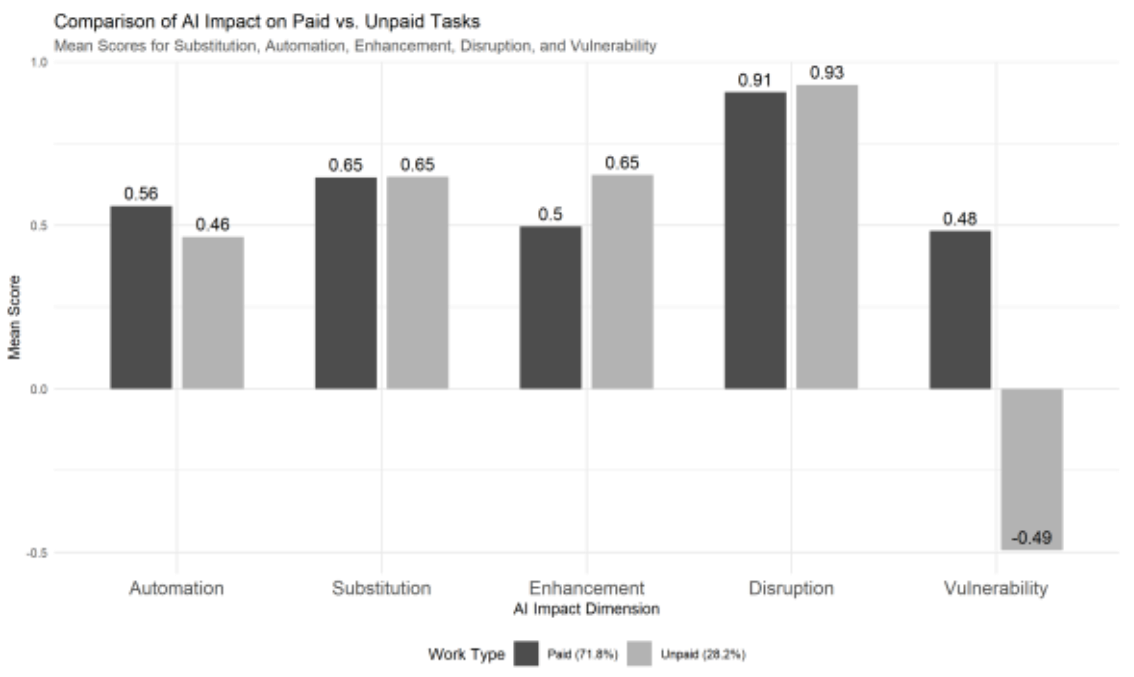

**Figure 5: Impact of AI on Platform Work: Paid vs Unpaid Tasks**

While the overall results suggest a somewhat similar offset effect in vulnerability and substitution scores, a closer examination at the occupational level reveals significant variation. Automation scores are consistently higher for paid work across all occupations, indicating that routine and repetitive tasks within paid labour are more susceptible to AI-driven automation. Freelance developers exhibit the highest automation score for paid work (0.72), followed by transcription (0.61), reflecting the structured nature of tasks in these fields (Figure 5.6).

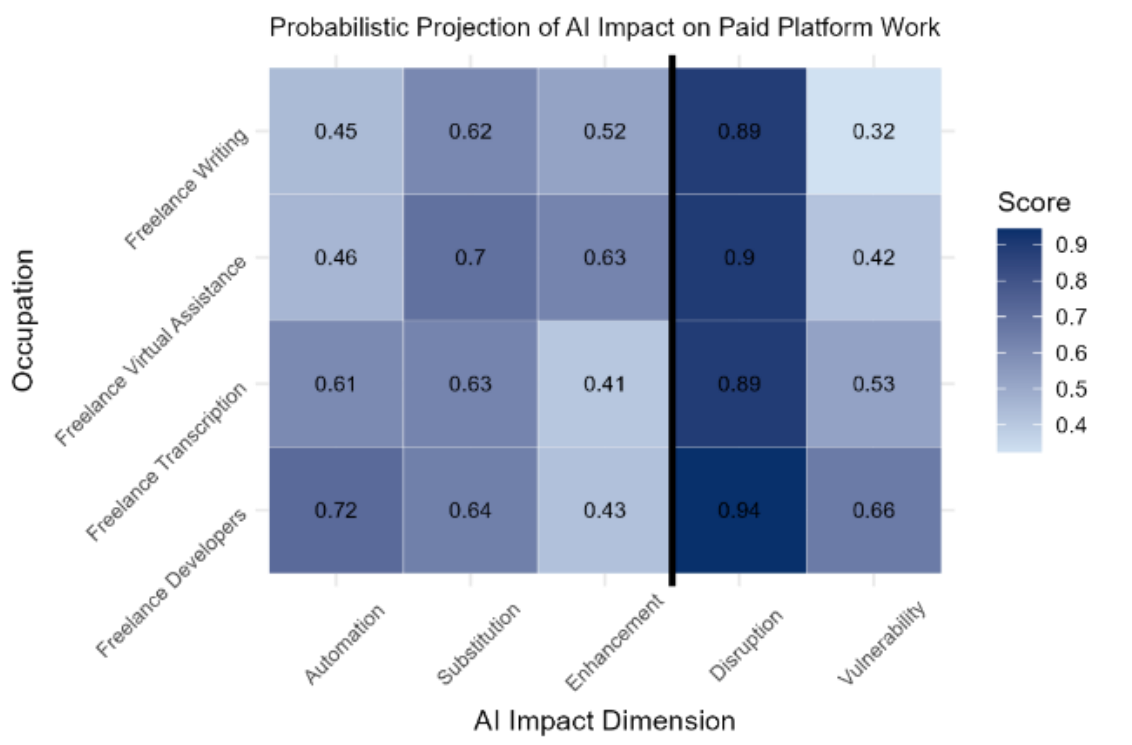

**Figure 6: AI Impact on Paid Work Tasks**

In contrast, automation scores for unpaid work are comparatively lower across all occupations, suggesting that non-remunerated tasks, often involving administrative, preparatory, or interpersonal components, are less likely to be replaced by AI (Figure 5.7). Among unpaid tasks, virtual assistance registers the lowest automation score (0.334), reinforcing the notion that tasks with lower economic value and requiring human discretion and adaptability remain resistant to automation.

Substitution effects reveal contrasting patterns across paid and unpaid work. In freelance development and transcription, unpaid tasks exhibit a higher potential for substitution than paid tasks (0.70 and 0.72 for unpaid work, compared to 0.64 and 0.63 for paid work, respectively). Conversely, virtual assistance and writing show the opposite trend, with unpaid substitution scores (0.59 and 0.57) slightly lower than their paid counterparts (0.70 and 0.62).

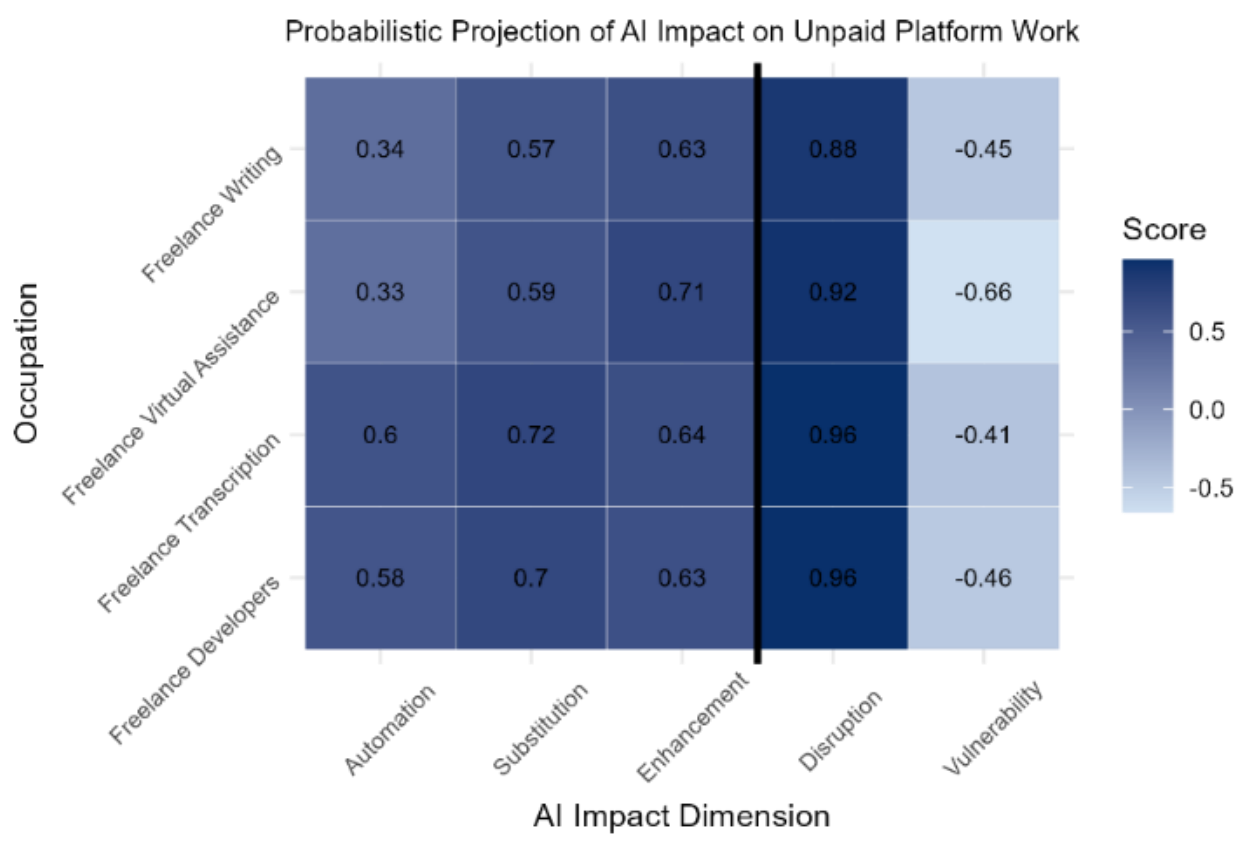

**Figure 7: AI Impact on Unpaid Tasks**

A notable trend across all observed platform-based occupations is the significantly higher enhancement scores for unpaid tasks compared to paid work, with an average difference of approximately 10 points. This suggests that AI-driven improvements are more likely to augment, rather than replace, the efficiency of tasks in proposal development, research, administrative, or coordination-related activities. Despite these variations, disruption scores remain consistently high across paid and unpaid work, suggesting that AI-driven changes profoundly affect both work categories, regardless of differences in automation, substitution, or enhancement effects.

A key finding in this comparison is the varying vulnerability scores across work categories. Unpaid tasks in freelance writing exhibit a vulnerability score of 0.45 (represented as -45 to indicate directionality), compared to 0.32 for paid tasks. This suggests that unpaid work within writing gigs benefits more from AI-driven transformations than paid tasks in the same occupation. Notably, freelance writing has the highest proportion of unpaid work (35.1%) among the occupations examined, contributing to a substantial offset effect. This is reflected in the overall livelihood vulnerability score (0.05) for freelance writing, indicating that the presence of unpaid tasks mitigates some of the adverse effects of AI on job security within this occupation.

Similarly, vulnerability scores for unpaid virtual assistance (0.66) and transcription work (0.41) are notably higher than paid tasks, which register scores of 0.42 and 0.53, respectively. The analyses in my research suggest that unpaid tasks in these occupations offset a portion of the impact of paid work vulnerability. As a result, the net vulnerability score for virtual assistance (0.21) and transcription work (0.23) remains more pronounced compared to freelance writing (0.05), where unpaid work plays a more substantial mitigating role.

Lastly, freelance developers exhibit a statistically higher vulnerability score in paid work (0.66) than in unpaid work (-0.46). Notably, the paid work vulnerability score for developers is the highest among all occupations examined, contributing to a significantly elevated net vulnerability score of 0.37. This highlights the heightened exposure of developers to AI-driven disruptions, particularly in revenue-generating coding tasks, where automation and substitution effects are more pronounced.

### Task-Level Impact Across Occupations

The key argument in this research is that AI's transformative impact is deeply intertwined with the intrinsic characteristics of tasks, shaping overall disruption and vulnerability at the occupational level. The analyses conform to the theoretical considerations that tasks with high modularity and extensive data availability, such as coding, proofreading, or drafting text, are particularly susceptible to automation and substitution when they do not require complex social context or a compensatory enhancement effect, resulting in high disruption and elevated vulnerability. Conversely, tasks demanding rich contextual understanding and significant social interaction, such as in-depth client consultations and communications, tend to resist automation. Their complexity and reliance on human judgment naturally mitigate some of the risks of automation. Additionally, the significance and nature of each task, reflected in its weight (W) and the proportion of paid work (R), further refine the impact of AI.

As detailed in the task-level results below, this integrated perspective, which links task characteristics such as modularity, context dependence, social complexity, data availability, and decision complexity to AI impact indices, offers a comprehensive framework for understanding how the nuanced nature of work influences occupation-level outcomes in digital platform labour.

#### *Freelance Writing*

Although the overall vulnerability score for freelance writing is relatively low (0.05), the interaction between task characteristics and AI impact reveals a more nuanced picture. Specific tasks within freelance writing, particularly those with high modularity and substantial data availability, such as proofreading and research, pose livelihood income risks for workers when they lack a compensatory enhancement effect. However, as evident in Table 5.2, a substantial proportion of freelance writing tasks require high contextual sensitivity and social interaction, providing greater resilience against automation and substitution risks.

**Table 2: Writing Tasks Characteristics Scoring**

| Freelance Writing Task Characteristics[8] | TMG | TCD | SIC | DAQ | DCU | $W_i$ | $R_i$ |
|---|---|---|---|---|---|---|---|
| Engage in initial client consultations to clarify project requirements, target audience, and desired tone or style. | 4 | 8 | 9 | 5 | 7 | 0.09 | 0.50 |
| Conduct comprehensive research on various topics tailored to the client's needs using credible sources to develop content ideas and detailed outlines tailored to the assignment. | 7 | 6 | 3 | 8 | 7 | 0.15 | 0.87 |
| Draft articles, blog posts, website content, academic papers, etc., adapting the writing style to suit various formats (creative, technical, academic, or copywriting). | 8 | 5 | 3 | 4 | 5 | 0.19 | 1.00 |
| Proofread the content thoroughly to eliminate grammatical, punctuation, and formatting errors, ensuring high-quality final output. | 8 | 2 | 1 | 9 | 3 | 0.12 | 0.87 |
| Regularly communicate with clients to provide progress updates, clarify requirements, and address feedback. | 7 | 4 | 6 | 7 | 7 | 0.08 | 0.71 |
| Revise and edit content based on client feedback, ensuring accuracy, tone, and adherence to guidelines. | 7 | 6 | 3 | 7 | 5 | 0.13 | 1.00 |
| Proactively search for and apply to new work opportunities by developing and submitting tailored proposals on freelance platforms and negotiating contracts, project terms, and rates with clients. | 7 | 5 | 6 | 7 | 6 | 0.12 | 0.00 |
| Manage administrative tasks such as invoicing, bookkeeping, and scheduling to sustain smooth business operations. | 9 | 3 | 3 | 9 | 3 | 0.04 | 0.00 |
| To attract potential clients, maintain and update an online portfolio and professional profile by regularly showcasing recent writing samples, client testimonials, and case studies. | 6 | 4 | 4 | 5 | 5 | 0.08 | 0.00 |

A closer look at task-level impact reveals significant variations in disruption and vulnerability. For instance, drafting, a core task with high weight (W = 0.19, Ri = 1), exhibits high modularity (TMG = 8) and moderate context dependence (TCD = 5). Its moderate data availability (DAQ = 4) results in a relatively low automation score (0.29) and a moderate substitution score (0.45). However, a high enhancement score (0.71) drives strong disruption (0.89) while keeping vulnerability low (0.09). This suggests that the structured nature of drafting work enables AI support without exposing workers to excessive risk.

[8] For readers' clarity, Wi represents the weight of the task, indicating its overall significance in the work process; and Ri is the ratio of paid work, distinguishing income-generating activities from unpaid functions. TMG (Task Modularity and Granularity), TCD (Context Dependence), SIC (Social & Interactive Complexity), DAQ (Data Availability & Quality), and DCU (Decision Complexity & Uncertainty). Annex D provides a comprehensive analysis to support the scoring presented in this table.

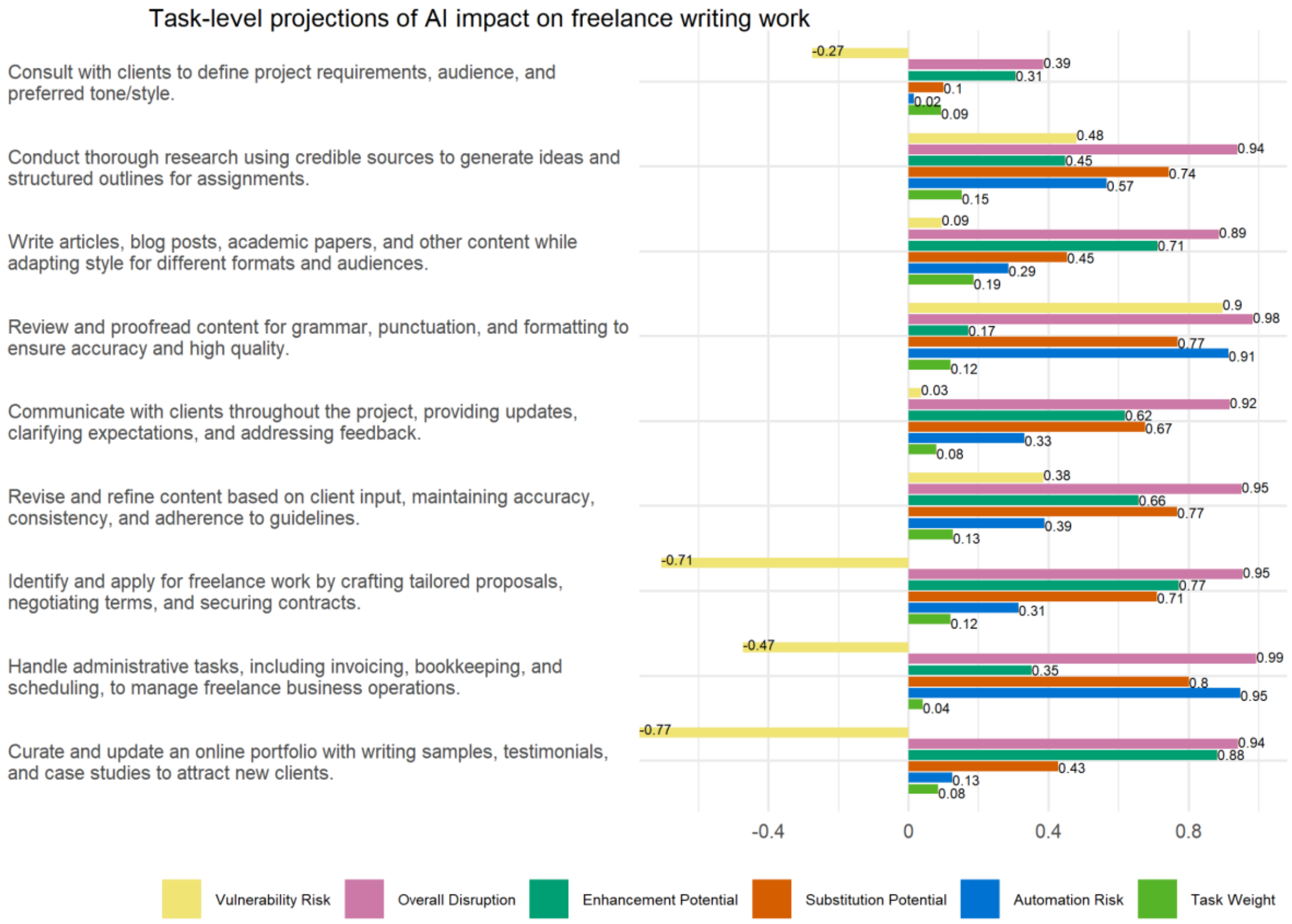

**Figure 8: Task-Level projections of AI impact Freelance Writing Work**

Contrastingly, proofreading (W = 0.12) demonstrates a markedly different risk profile. Despite sharing high modularity (TMG = 8) with drafting, proofreading benefits from very high data availability (DAQ = 9) but has minimal social complexity (SIC = 1) and low context dependence (TCD = 2). As a result, it becomes highly susceptible to automation (0.91) and substitution (0.77). Furthermore, limited enhancement potential (0.17) fails to counterbalance these risks, leading to near-maximal disruption (0.98) and high vulnerability (0.90). This illustrates how even structurally similar tasks can diverge in risk when enhancement effects are absent (Figure 5.8).

In contrast, client consultations exhibit a distinctly protective profile. With low modularity (TMG = 4) but very high context dependence (TCD = 8) and social complexity (SIC = 9), this task naturally resists automation (0.02) and substitution (0.10). Modest enhancement (0.31), coupled with its intrinsic interpersonal demands, results in low disruption (0.39) and even negative vulnerability (-0.27), indicating a protective effect partly due to its mixed paid ratio (Ri = 0.50).

Similarly, research work, characterized by high modularity (TMG = 7) and robust data availability (DAQ = 8) yet lower social complexity (SIC = 3) and moderate context dependence (TCD = 6), exhibits elevated substitution (0.74) and automation (0.57) scores. These factors drive near-maximal disruption (0.94) and moderately high vulnerability (0.48), particularly due to its significant weight (W = 0.15) and high paid ratio (Ri = 0.87). In contrast, client communication, which involves moderate modularity (TMG = 7), modest context dependence (TCD = 4), and higher social complexity (SIC = 6), maintains a low automation potential (0.33) while benefiting from a strong enhancement score (0.62). This combination results in high disruption (0.92) yet minimal vulnerability (0.03), underscoring the buffering effect of interpersonal interaction.

These findings, shown in Figure 5.8, illustrate that while freelance writing as an occupation remains relatively insulated from AI-driven vulnerability, task-level differences reveal significant disparities in exposure. Highly modular, data-rich tasks like proofreading and research are disproportionately susceptible to automation and substitution. In contrast, tasks requiring human judgment, social interaction, and contextual understanding, such as client consultations and communication, are more resistant. Notably, the degree of enhancement plays a crucial role in determining whether disruption translates into vulnerability. These dynamics suggest that AI's role in freelance writing will likely be augmentation rather than wholesale replacement, with human expertise remaining a central component in many core tasks.

#### *Freelance Virtual Assistance*

At the occupational level, freelance virtual assistance exhibits the highest substitution score (0.61) and enhancement score (0.64) among the examined occupations. These trends are closely linked to the underlying characteristics of tasks within this role (Table 5.3 and Figure 5.9). While client-facing and decision-oriented tasks (e.g., consultations, project management) demonstrate greater resistance to automation, structured and repetitive tasks (e.g., scheduling, correspondence, and administrative duties) are highly vulnerable to substitution.

For instance, initial client consultations, which involve direct engagement with clients, display low modularity (TMG = 4) but high context dependence (TCD = 8) and strong social complexity (SIC = 9). These features naturally shield the task from automation and substitution, as reflected in its low automation (0.02) and substitution (0.10) scores. Consequently, it experiences only modest disruption (0.39) and even negative vulnerability (-0.15) due to its relatively balanced paid work ratio (Ri = 0.87).

**Table 3: Freelance Virtual Assistance Task Characteristics Scoring**

| Freelance Virtual Assistance | TMG | TCD | SIC | DAQ | DCU | $W_i$ | $R_i$ |
|---|---|---|---|---|---|---|---|
| Engage in initial client consultations to clarify project requirements, communication expectations, and preferred tools, establishing clear roles for long-term or short-term engagements. | 4 | 8 | 9 | 5 | 7 | 0.09 | 0.87 |
| Manage and organize schedules and appointments using digital calendaring and scheduling software to ensure optimal time management and timely coordination of meetings or tasks. | 7 | 5 | 4 | 8 | 6 | 0.10 | 1.00 |
| Handle email correspondence and other communications, liaising between clients and their contacts by filtering, prioritizing, and responding to messages as needed. | 8 | 6 | 4 | 8 | 7 | 0.11 | 1.00 |
| Conduct targeted research and compile data to prepare reports, presentations, or summaries that support the client's decision-making and project objectives. | 7 | 7 | 4 | 7 | 7 | 0.11 | 1.00 |
| Assist with project management by coordinating tasks, tracking progress using project management software, and ensuring deadlines and milestones are met. | 7 | 6 | 5 | 8 | 6 | 0.11 | 1.00 |
| Manage social media accounts by scheduling content, monitoring engagement, and performing essential updates per the client's brand guidelines. | 7 | 6 | 7 | 8 | 6 | 0.11 | 1.00 |
| Perform administrative tasks, such as document preparation, data entry, travel arrangements, and running errands, while coordinating with other freelancers or team members to ensure seamless execution of projects. | 8 | 5 | 5 | 9 | 4 | 0.10 | 1.00 |
| Maintain regular and proactive communication with clients by providing progress updates, clarifying requirements, and promptly addressing any concerns or emerging issues. | 7 | 6 | 6 | 7 | 7 | 0.10 | 1.00 |
| Actively search for new work opportunities by researching freelance platforms, developing and submitting tailored proposals, and negotiating contracts, project terms, and rates with clients. | 7 | 5 | 6 | 7 | 6 | 0.09 | 0.00 |

| Freelance Virtual Assistance | TMG | TCD | SIC | DAQ | DCU | $W_i$ | $R_i$ |
|---|---|---|---|---|---|---|---|
| Manage financial and administrative systems by tracking invoices, processing payments, and maintaining tax documentation to support smooth business operations. | 9 | 3 | 3 | 9 | 3 | 0.03 | 0.00 |
| Maintain and update an online portfolio and professional profile by regularly showcasing recent work. | 6 | 4 | 4 | 5 | 5 | 0.07 | 0.00 |

By contrast, routine digital organization tasks, such as scheduling and appointment management, exhibit high modularity (TMG = 7) and strong data availability (DAQ = 8) but only moderate context dependence (TCD = 5) and low social complexity (SIC = 4). These characteristics make such tasks highly susceptible to substitution (0.84) and moderately automatable (0.54). However, despite a strong enhancement effect (0.74), a moderate vulnerability level (0.50) highlights how structured, repetitive tasks are at risk despite AI-driven efficiency gains.

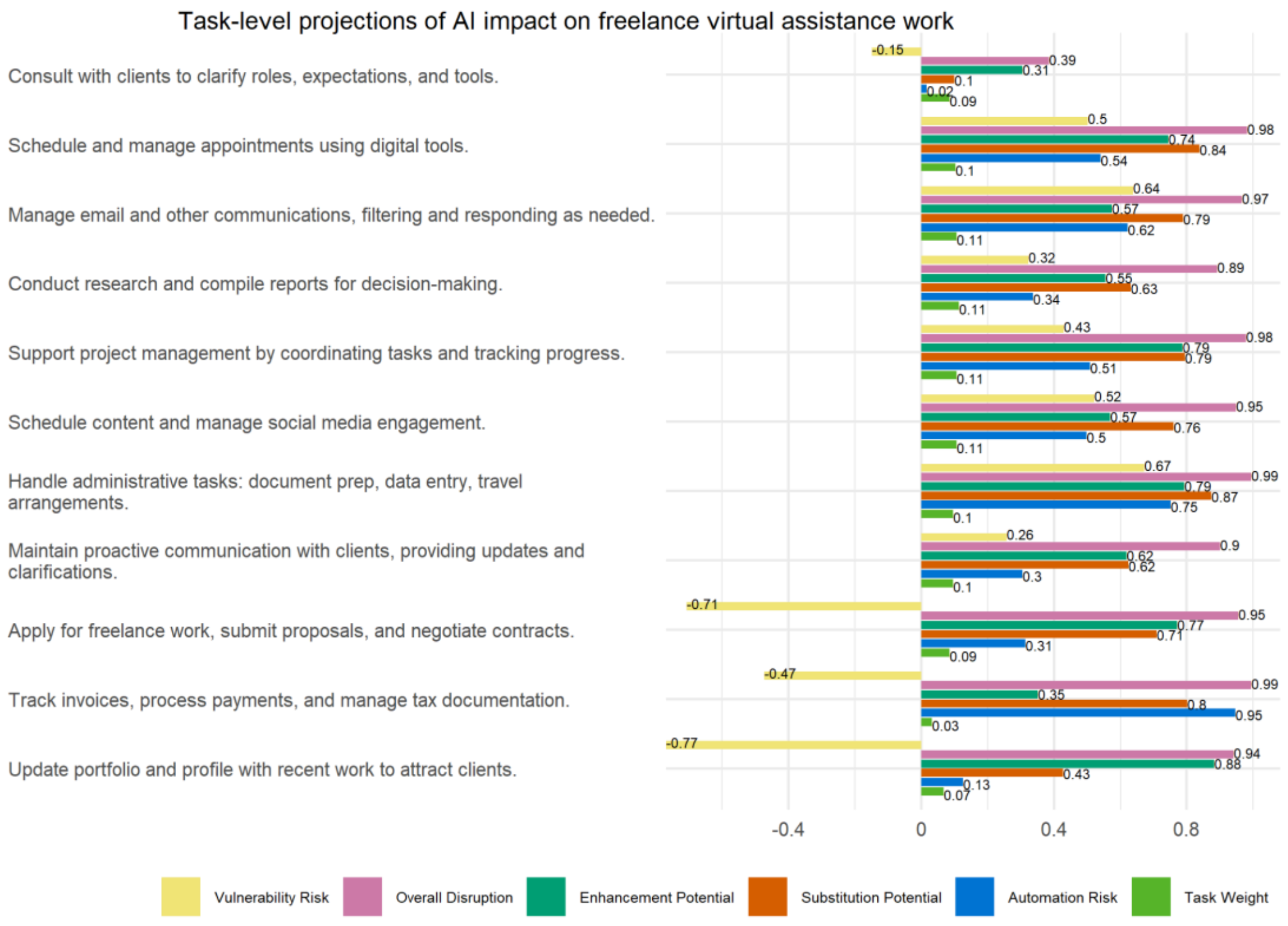

**Figure 9: Task-level projections of AI impact on Freelance Virtual Assistance**

Similarly, handling correspondence and communication shares a high modularity score (TMG = 8) and strong data availability (DAQ = 8) but only moderate context dependence (TCD = 6) and low social complexity (SIC = 4). As a result, it faces significant automation (0.62) and high substitution potential (0.79). Its moderate enhancement score (0.57) is insufficient to fully counterbalance these pressures, leading to high disruption (0.97) and elevated vulnerability (0.64).

Conversely, project management tasks, which involve tracking progress and ensuring a timely execution of projects, demonstrate a more balanced risk profile. With moderate modularity (TMG = 7), moderate context dependence (TCD = 6), and moderate social complexity (SIC = 5), project management tasks benefit from robust data availability (DAQ = 8). These attributes contribute to very high substitution (0.79) and moderate automation (0.51), yet high enhancement (0.79) helps offset some risks. While the overall disruption score remains exceptionally high (0.98) due to the magnitude of change, vulnerability is also significant (0.43) because of the heightened substitution risk, which may lead to skill degradation and financial instability.

Finally, administrative tasks, while essential for operational efficiency, are among the most vulnerable due to their highly routine nature. These tasks score high in modularity (TMG = 8) and data availability (DAQ = 9) but exhibit low context dependence (TCD = 5) and moderate social complexity (SIC = 5). As a result, they record high substitution (0.87) and automation (0.75) scores. Even with a strong enhancement effect (0.79), the nearly maximal disruption (0.99) leads to high vulnerability (0.67), making administrative functions one of the most at-risk categories in virtual assistance work.

#### *Transcription Services*

Transcription work is shaped by strict accuracy requirements and adherence to client guidelines, which influence its exposure to AI-driven changes. The core task of converting audio or video recordings into text is fully remunerated (Ri = 1.00, W = 0.16). It involves a high degree of context dependence (TCD = 8) and considerable decision complexity (DCU = 7) due to the need for precise transcription with minimal errors (Table 5.4). These characteristics result in moderate automation (0.24) and high substitution (0.40) scores, with modest enhancement potential (0.35). Consequently, this task registers a moderately low vulnerability (0.23), indicating that while AI may aid transcription, full automation remains constrained by the necessity for accuracy and compliance with client specifications.

**Table 4: Transcription Service Task Characteristics Scoring**

| Freelance Transcription Service | TMG | TCD | SIC | DAQ | DCU | $W_i$ | $R_i$ |
|---|---|---|---|---|---|---|---|
| Convert audio or video recordings into text following client formatting guidelines, including any required timestamps and speaker identifications. | 6 | 8 | 2 | 5 | 7 | 0.16 | 1.00 |
| Research and integrate domain-specific terminology and contextual details, especially for technical or specialized content. | 7 | 8 | 3 | 8 | 8 | 0.14 | 0.87 |
| Rigorously review transcripts to correct errors in grammar, punctuation, and spelling. | 8 | 2 | 1 | 9 | 3 | 0.15 | 1.00 |
| Consult with clients to clarify unclear or ambiguous audio segments, ensuring the final transcript accurately reflects the original content. | 6 | 7 | 6 | 7 | 7 | 0.12 | 0.71 |
| Update and refine transcripts in response to client corrections and detailed feedback. | 7 | 6 | 3 | 7 | 5 | 0.13 | 1.00 |
| Provide regular updates and clarifications during transcription to align the final output with client expectations. | 6 | 5 | 4 | 7 | 5 | 0.07 | 0.50 |
| Proactively search for and submit tailored proposals on freelance platforms while negotiating client contracts, project terms, and rates. | 7 | 5 | 4 | 7 | 6 | 0.10 | 0.00 |
| Manage administrative tasks such as invoicing, bookkeeping, and scheduling to sustain smooth business operations. | 9 | 3 | 2 | 9 | 3 | 0.05 | 0.00 |
| Maintain and update your professional profile. | 6 | 4 | 3 | 5 | 5 | 0.08 | 0.00 |

A notable contrast emerges when examining proofreading and transcript review, both of which are crucial tasks for ensuring quality. It exhibits high modularity (TMG = 8) and very high data availability (DAQ = 9) but minimal social complexity (SIC = 1) and low context dependence (TCD = 2). These characteristics make it highly susceptible to automation (0.84) and substitution (0.54). Furthermore, limited enhancement potential (0.17) fails to counterbalance these risks, resulting in near-maximal disruption (0.98) and high vulnerability (0.90). This highlights that even tasks integral to transcription can become increasingly at risk when they lack contextual or interpersonal complexity.

Similarly, research tasks, which involve integrating domain-specific terminology and contextual details, exhibit high modularity (TMG = 7), high context dependence (TCD = 8), and slightly elevated social complexity (SIC = 3) due to the specialized knowledge required for industry-specific transcripts. These characteristics, coupled with robust data availability (DAQ = 8) and high decision complexity (DCU = 8), lead to high automation (0.77) and substitution (0.63) scores. As a result, this task experiences significant disruption (0.94) and high vulnerability (0.68), illustrating that even knowledge-intensive transcription tasks remain highly susceptible to AI replacement.

A different pattern emerges in peripheral work, such as consultation with clients to clarify ambiguous audio segments, which, although performed on a needs basis (W = 0.12, Ri = 0.71), relies on moderate modularity (TMG = 6), high context dependence (TCD = 7), and moderate social complexity (SIC = 6). These attributes yield some automation potential (0.51) and high substitution (0.70). However, strong enhancement (0.58) helps mitigate some of these risks, resulting in high disruption (0.94) but comparatively lower vulnerability (0.17). This underscores how tasks that incorporate interpersonal judgment remain relatively resilient to full AI displacement.

Similarly, routine revision tasks, such as updating transcripts based on client feedback, a fully paid task (Ri = 1.00, W = 0.13), exhibit moderate modularity (TMG = 7), context dependence (TCD = 6), and social complexity (SIC = 3). However, high automation (0.73) and substitution (0.90) scores indicate that AI can significantly streamline revision processes. Despite a moderate enhancement potential (0.66), the rapid pace of AI-driven transcription correction leaves workers vulnerable, resulting in a high vulnerability score (0.74). This suggests that even tasks benefiting from enhancement remain at considerable risk due to AI's growing efficiency in transcription-based revisions.

Overall, the task-level analysis of transcription work that many platform workers do shows contrasting impacts of AI, depending on the nature of the tasks. While core transcription activities exhibit moderate automation and substitution risks, tasks with high modularity and robust data availability, such as proofreading and transcript revisions, face substantially higher displacement risks due to their repetitive and structured nature (Figure 5.10).

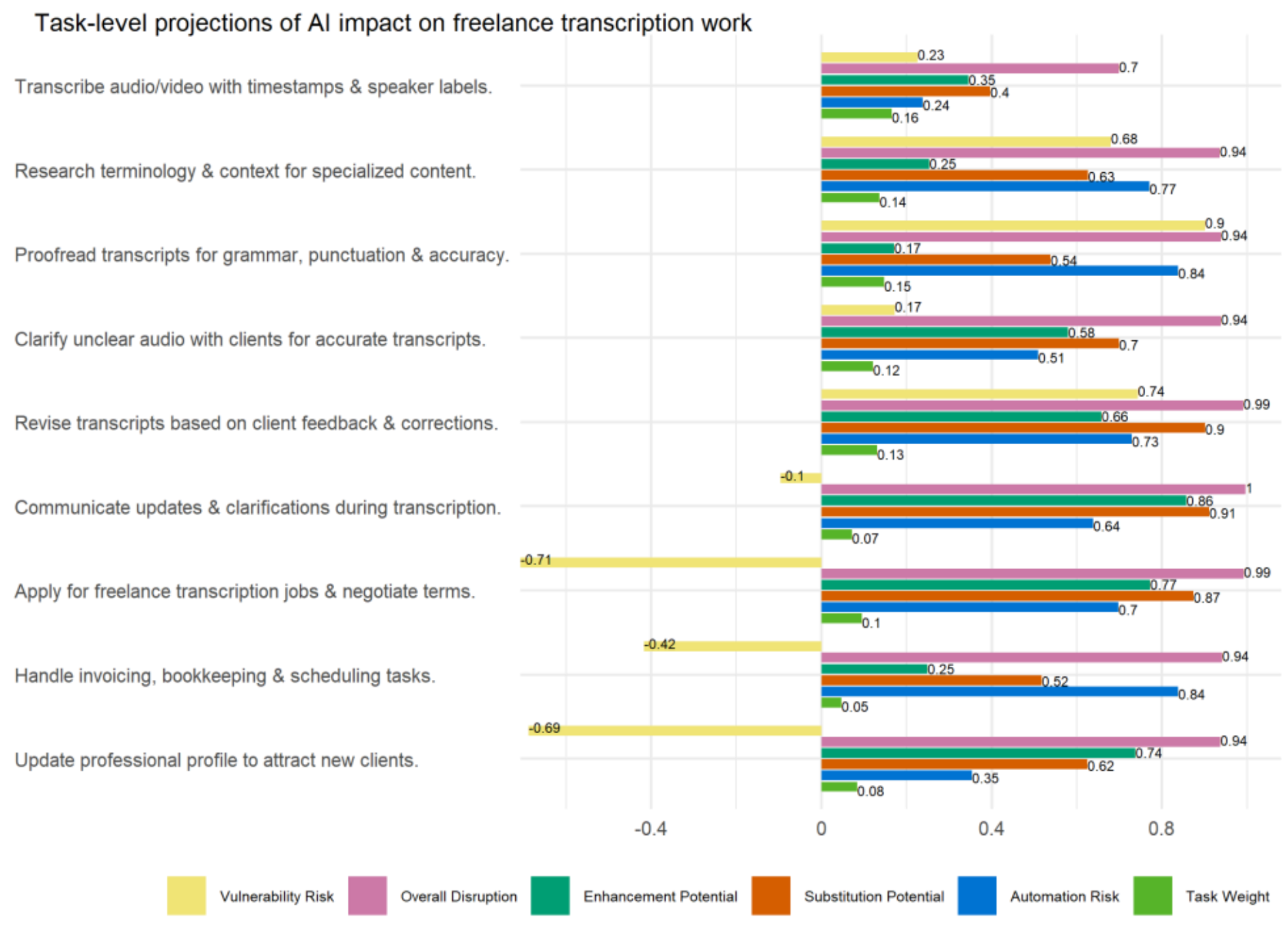

**Figure 10: Task-level projections of AI impact on Transcription Work**

However, tasks requiring interpersonal communication and domain-specific expertise, such as client consultations and research-based transcription, remain more resilient due to their contextual dependencies and specialized decision-making processes. Nonetheless, the overall vulnerability of transcription work remains high, particularly for tasks that AI can rapidly enhance but also replace, leading to heightened job insecurity for freelancers in this occupation.

#### *Freelance Developers*

Among the occupations examined, freelance software and web development stand out with the highest automation score (0.68) and vulnerability score (0.37), reflecting the diverse nature of tasks and their susceptibility to AI-driven transformation (Table 5.5). The highly structured nature of core technical tasks, such as coding and software maintenance, makes them particularly exposed to automation.

For instance, developing and maintaining code, a fundamental core paid task (Ri = 1.00, W = 0.14), exhibits high modularity (TMG = 8) and robust data availability (DAQ = 8), while displaying low social complexity (SIC = 3) and moderate context dependence (TCD = 6). These characteristics predispose coding to high automation (0.83) and moderate substitution (0.64). Despite some enhancement potential (0.43), the resulting near-maximal disruption (0.97) leads to high vulnerability (0.81)—a risk further intensified by its fully paid nature and substantial task weight.

A similar pattern emerges in tasks such as website and application optimization, where the interplay of high modularity (TMG = 7), strong data availability (DAQ = 8), and elevated context dependence (TCD = 8) leads to substantial automation (0.75) and moderate enhancement (0.40). This combination results in high disruption (0.95) and considerable vulnerability (0.76), illustrating how even tasks requiring some degree of context adaptation remain highly susceptible to AI-driven changes.

**Table 5: Web and Software Development Task Characteristics Scoring**

| Freelance Developers | TMG | TCD | SIC | DAQ | DCU | $W_i$ | $R_i$ |
|---|---|---|---|---|---|---|---|
| Communicate with clients to clarify project requirements, desired functionalities, and target objectives, ensuring a clear understanding of the assignment. | 6 | 7 | 6 | 7 | 7 | 0.10 | 0.50 |
| Conduct technical research to identify effective solutions, best practices, and appropriate technologies to inform the development process. | 7 | 8 | 3 | 8 | 8 | 0.12 | 0.87 |
| Develop and maintain code to meet specific client requirements related to custom software and web applications. | 8 | 6 | 3 | 8 | 7 | 0.14 | 1.00 |
| Optimize websites, content management systems (e.g., WordPress, Drupal), and applications for performance, speed, security, and reliability as needed or per client request. | 7 | 8 | 4 | 8 | 8 | 0.12 | 1.00 |
| Document code and technical processes to ensure quality and maintainability. | 8 | 5 | 3 | 9 | 5 | 0.10 | 1.00 |
| Troubleshoot and resolve technical issues in existing systems and codebases, including debugging and addressing production challenges. | 6 | 8 | 3 | 7 | 9 | 0.13 | 1.00 |
| Provide ongoing maintenance, updates, and improvements to ensure the long-term viability and scalability of the project. | 6 | 7 | 4 | 8 | 7 | 0.09 | 1.00 |
| Proactively search for and apply to new work opportunities by developing and submitting tailored proposals on freelance platforms and negotiating contracts, project terms, and rates with clients. | 7 | 5 | 6 | 7 | 6 | 0.10 | 0.00 |
| Manage administrative tasks such as invoicing, bookkeeping, and scheduling to sustain smooth business operations. | 9 | 3 | 2 | 9 | 3 | 0.03 | 0.00 |
| To attract potential clients, maintain and update an online portfolio and professional profile by regularly showcasing recent work samples, client testimonials, and case studies. | 6 | 4 | 3 | 5 | 5 | 0.07 | 0.00 |

In contrast to the high automation risks observed in technical tasks, those involving direct client communication, such as clarifying project requirements, demonstrate greater resilience (but less income-generating value). With moderate modularity (TMG = 6), high context dependence (TCD = 7), and balanced social complexity (SIC = 6), this task is less susceptible to automation (0.51) and substitution (0.70). These factors, combined with a moderate enhancement potential (0.58), result in high disruption (0.94) but near-neutral vulnerability (-0.05), primarily because this task is only partially paid (Ri = 0.50).

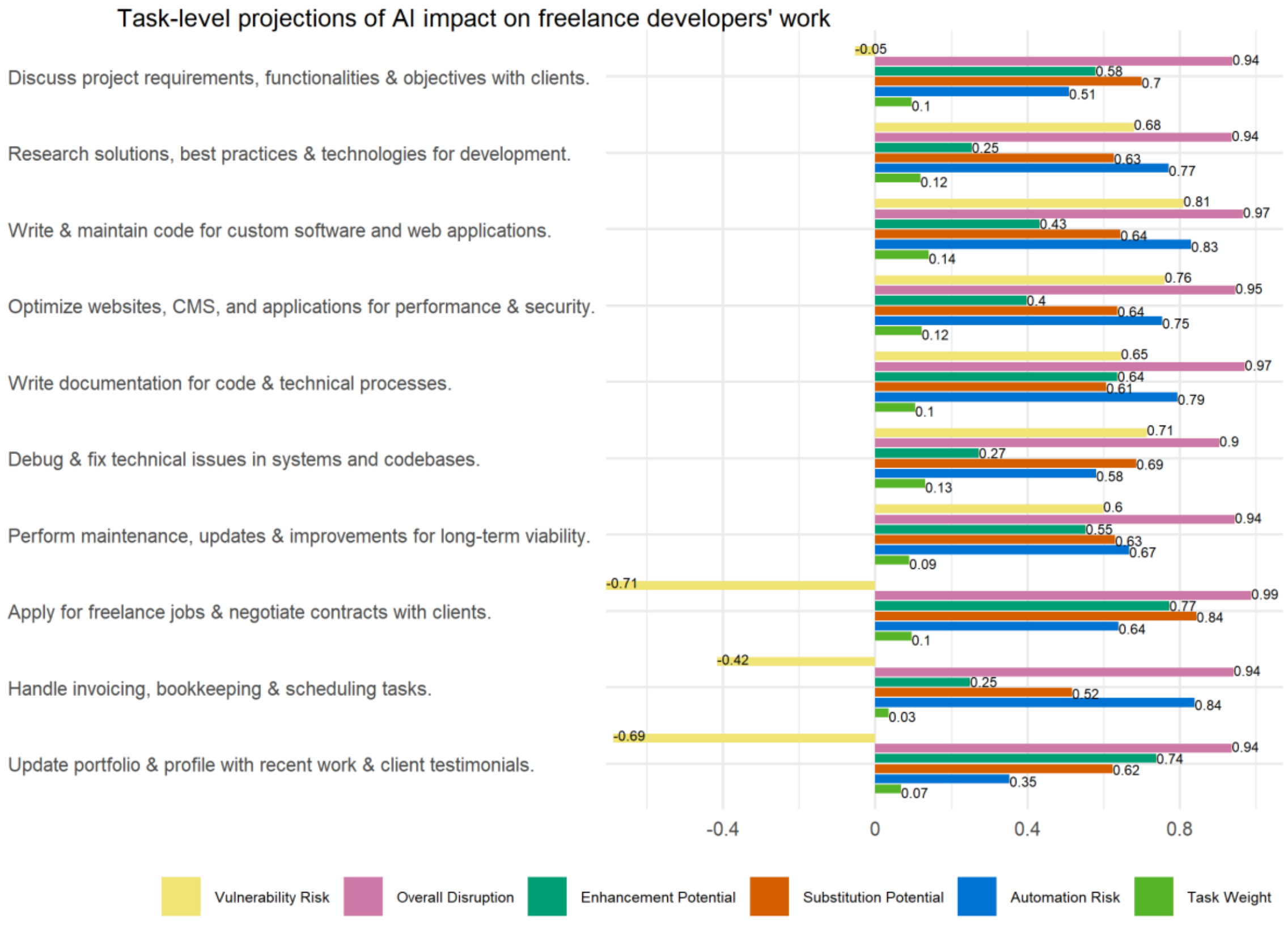

**Figure 11: Task-Level projections of AI impact on Web and Software Development Work**

Other structured but essential tasks, such as documenting code, present a unique case. This task benefits from very high data availability (DAQ = 9) and high modularity (TMG = 8), while maintaining moderate context dependence (TCD = 5) and low social complexity (SIC = 3). Here, moderate automation (0.79) combined with a more substantial enhancement effect (0.64) lowers vulnerability to moderate levels (0.65), even though disruption remains near-maximal (0.97).

Meanwhile, troubleshooting technical issues exemplifies the complex trade-offs in AI's impact on development work. With moderate modularity (TMG = 6), high decision complexity (DCU = 9), and strong context dependence (TCD = 8), this task faces moderate automation (0.58) and substitution (0.69) scores. However, its limited enhancement potential (0.27) results in high disruption (0.90) and elevated vulnerability (0.71). This suggests that while troubleshooting remains a critical function, AI-driven diagnostics and debugging tools are increasingly encroaching on aspects of this work.

Unlike other occupations examined, freelance development work presents a paradoxical relationship with AI. The high modularity, strong data availability, and structured nature of technical tasks make them prime candidates for automation, leading to significant disruption and vulnerability (Figure 5.11). However, tasks requiring decision complexity, adaptation, and client interaction remain somewhat insulated from full automation, though they still face substantial shifts in workflow and skill demands.

## *Discussion and Empirical Synthesis*

In this research, I thoroughly examined the task characteristics of platform work to assess the potential impact of generative AI on the nature of available work and the livelihoods of platform workers. The results in this study particularly highlight the significant differences between platform-based labour and conventional employment by demonstrating how key work tasks across various platform work categories differ due to the nature of work available and the ratio of unpaid tasks. Analyzing these differences offers valuable insights into how professional identities and job titles in the digital economy may mirror those in traditional labour markets; however, the structure of available work and the burden of unpaid labour create nuanced variations in workers' exposure to transformations in the labour market.

While occupations such as web development, personal assistance, and content writing may seem similar to traditional professional identities, their nature within the platform economy is fundamentally different. In the platform economy, these professions heavily depend on hyper-digitalized, fragmented, and data-driven job tasks that enable firms to profit from precarious labour while making low-commitment gigs attractive to workers. However, the precariousness of platform-based work extends beyond inconsistent job availability and fluctuating wages; it reflects a structural constraint within the labour market. The very factors that enable task fragmentation, letting firms and freelancers operate within this ecosystem of precarious work, also increase workers' exposure to market volatility and technological disruptions.

### Contextualizing Uneven AI Disruption and Its Perceived Impact

As the findings of the theoretical analysis suggest, AI-driven automation in data-intensive tasks could result in considerable direct replacement of labour with capital (AI). Even in cases where the social and decision complexity of tasks decreases the likelihood of full automation, the partial substitution of human labour may have a dual impact. While it could enhance productivity, it may also introduce particular livelihood vulnerabilities by suppressing wages and diminishing the skill premium.

Several key practical questions arise from the theoretical projections of AI's impact on categories of platform work. For instance, if the livelihood vulnerability (0.214) associated with generative AI is expected to suppress roughly a fifth of platform work, and the mean disruption score of 0.913 (both on a scale of 0 to 1) indicates a significantly high level of disruption, what observable early-stage changes in the job market might emerge at this stage? Moreover, how does this transformation align with the adoption of AI tools? Some studies estimate a 24% decline in job supply on a major digital platform within two years of ChatGPT's introduction (Liu et al., 2023). This demand-side contraction, representing approximately a quarter of platform jobs, along with the theoretical inference from my analysis that AI-induced livelihood vulnerability could be of a similar magnitude, suggests that occupations most susceptible to AI disruption are already facing a sharp decline in job availability.

However, to what extent is this early-stage transformation evident in the perceptions of workers? To explore this question, I conducted field research with Kenyan platform workers from January 2024 to September 2024, which included examining their views on the impact of AI on their livelihood. Their responses, however, reveal a different story.

**Table 6: Perceived Job Insecurity Due to AI Among Freelance Platform Workers (N=41)**

| Occupation | Mean | SD | Min | Max |
|---|---|---|---|---|
| Freelance Writing | 2.44 | 1.15 | 0 | 4 |
| Transcription | 1.28 | 0.95 | 0 | 3 |
| Virtual Assistance | 1.76 | 1.30 | 0 | 3 |
| Web and Software Development | 2.00 | 1.41 | 0 | 4 |

0 = not at all concerned, 4 = extremely concerned

As shown in Table 5.6, survey responses from my research indicate that freelancers engaged in writing gigs, who, according to these theoretical projections, face the lowest livelihood vulnerability (0.05), express the highest level of concern about AI's impact on their future work opportunities [9]. Web and software development workers follow as the second most concerned group. However, the overall concern among workers remains moderate, which is not expected if a quarter of their livelihood opportunities have already disappeared. This does not necessarily contradict the observed decline in job availability or the broader impact of AI but suggests that additional factors may influence workers' perceptions and concerns.

[9] Please note that the timing of the survey is important for interpreting the findings. The data were collected between January and July 2024, and it is possible that workers' views have since evolved.

First, market fluctuations likely affect different groups of workers unevenly. Those who have established a clientele and a surplus of work have more control over their opportunities, making them less susceptible to early-stage disruption. It should be noted that the platform labour market is widely recognized as oversupplied, with significantly more workers seeking opportunities than available job postings (Kässi et al., 2021). A conventional labour supply perspective, focusing primarily on the job-to-worker ratio, may overlook a crucial aspect of platform work: the signalling power of workers' profiles. The established trust, reinforced by indirect network effects and review-based credibility, is critical in shaping employment outcomes and the amount of work available to workers.

More specifically, the exponential growth of many digital platforms with a two-sided market is significantly influenced by indirect network effects and the pivotal role of ratings and reviews in user decision-making, resulting in a winner-takes-most scenario for both platforms and workers. Indirect network effects occur when the value of a platform increases for one user group as the size of another user group expands (Lee & Wellman, 2012). For instance, increasing the number of work providers in digital labour platforms leads to more job opportunities, attracting more freelancers. Conversely, a larger pool of freelancers offers clients a broader selection of skills, enhancing the platform's appeal to employers. This reciprocal relationship fosters a self-reinforcing cycle of growth, conveying a sense of confidence to both sides of the market that their business needs will be met.

Sustaining this growth and user engagement, however, depends on mechanisms that demonstrate continued trust. One of the primary mechanisms for building this trust in the platform economy is through ratings and reviews. Empirical studies have shown that online reviews have a significant impact on key platform decisions, such as purchasing a product or hiring a worker (Pulignano et al., 2025; Yoganarasimhan, 2013). This creates a feedback loop where well-reviewed freelancers secure more projects, leading to greater control over their work options and additional reviews that further enhance their attractiveness to potential clients.

As a result, freelancers with stronger signalling power are more likely to attract a surplus of clients and have greater confidence in their ability to sustain adequate work access under disruptive conditions, as well as find work on favourable terms in the future. Unlike traditional employment and many location-based work platforms, a worker on a digital labour platform does not appear busy, engaged in a task, or employed full-time. Thus, a well-positioned worker could theoretically accumulate a significant amount of work. In some instances, a well-placed individual profile could technically gather enough work for a whole team of workers (Idowu & Elbanna, 2021).

In the survey, I asked workers about their confidence in finding sufficient work through digital platforms over the next six months. Table 5.7 displays the results of a linear regression model that examines relevant factors.

**Table 7: Determinants of Workers' Confidence in Future Digital Platform Employment[10]**

| Work Access Confidence | b | SE | p | 95% confidence interval of *b* | | beta |
|---|---|---|---|---|---|---|
| | | | | Lower | Upper | |
| AI Job Loss Concern | -0.103 | 0.075 | 0.174 | -0.254 | 0.047 | -0.174 |
| Paid Work Hours | 0.0005 | 0.005 | 0.925 | -0.01 | 0.009 | -0.012 |
| Age | 0.089 | 0.136 | 0.519 | -0.186 | 0.363 | 0.094 |
| Education Level | 0.057 | 0.174 | 0.743 | -0.294 | 0.409 | 0.043 |
| Gig Experience | 0.037 | 0.13 | 0.780 | -0.226 | 0.3 | 0.039 |

[10] The model explains 40.35% of the variance in confidence to access work ($R^2 = 0.4035$, Adjusted $R^2 = 0.2899$), with the overall model being statistically significant ($F(8, 42) = 3.55$, $p = 0.0032$). Pre-estimation checks confirmed that multicollinearity is not a concern (VIF < 5 for all predictors). The Shapiro-Wilk test ($p = 0.556$) indicates that residuals are normally distributed. The Breusch-Pagan test ($p = 0.7114$) found no evidence of heteroskedasticity, supporting the use of standard errors. The Ramsey RESET test ($p = 0.3404$) suggests the model does not suffer from omitted variable bias.

| | | | | | | |
|---|---|---|---|---|---|---|
| Work Control | 0.275* | 0.074 | 0.001 | 0.127 | 0.424 | 0.475 |
| Financial Satisfaction | 0.35* | 0.11 | 0.003 | 0.128 | 0.572 | 0.414 |
| Gender | -0.182 | 0.178 | 0.312 | -0.542 | 0.177 | -0.127 |
| Intercept | 1.141 | 0.808 | 0.165 | -0.49 | 2.772 | - |

n=51
*p<.05

Among the predictors listed in Table 5.7, control over workload, schedule, and financial satisfaction emerge as the only significant factors influencing workers' confidence in securing future platform jobs. Work control (B = 0.275, SE = 0.074, p = 0.001, β = 0.475) has the strongest positive effect, demonstrating that workers who feel they have greater autonomy over their tasks, schedules, and job selection exhibit significantly higher confidence in their ability to secure future digital platform work. These findings suggest that platform workers with greater control over their workflow perceive their employment as more stable and predictable, thereby reducing their vulnerability to job insecurity and fluctuations in demand.

Likewise, workers' negotiating power leading to financial satisfaction (B = 0.350, SE = 0.110, p = 0.003, β = 0.414) is positively linked to confidence in continuing to find suitable work. This suggests that financially secure workers tend to be more optimistic about their ability to keep earning through digital platforms in the future, irrespective of the ongoing transformation. This finding is consistent with broader labour market trends, where financial stability is a buffer against economic uncertainty. In the context of platform work, it may suggest that workers who have attained financial security by enhancing their platform profiles view their income streams as reliable and replicable. In contrast, those experiencing financial instability and difficulties attracting sufficient clients might perceive more significant risks in sustaining their platform-based employment, especially if automation and substitution reduce available work opportunities.

Other variables, such as hours worked, age, education level, total experience, and gender, do not show significant effects on workers' confidence in securing future platform jobs (p > 0.05 for all predictors). The findings suggest that demographic factors and work history alone do not dictate confidence in maintaining digital platform work. It is particularly noteworthy that gig experience does not show a significant impact, indicating that time spent on platform work does not necessarily lead to increased confidence in future opportunities. This reflects the precarious nature of digital labour, where even experienced workers may still encounter instability and uncertainty in obtaining jobs if their years of experience do not result in greater control over platform work conditions and bargaining power.

Importantly, concerns regarding AI-related job loss (B = -0.103, SE = 0.075, p = 0.174, β = -0.174) do not have a statistically significant impact, indicating that worries about automation and AI replacing jobs on digital platforms exist but do not currently align with workers' confidence in their ability to secure future employment. However, it is possible that as automation advances, this effect could become more pronounced over time. My research indicates that workers' confidence in their ability to secure future employment is also a function of their resilience in the face of persistent labour market precariousness. The inherent instability of platform work conditions them to adapt to emerging transformations. Theo, one of the research participants I spoke to, explained this:

> "Freelancing is very tricky because you don't have a regular schedule. So today, you may have work, a lot of work, then in the next three months there are no jobs coming your way. So, let me say it's very unpredictable (Theo, software and web developer)"

Theo started freelancing during his undergraduate studies; for him, unpredictable workloads and finding work through platforms amid uncertainty have become defining features of his career, particularly after graduating in 2020 when COVID-19 restrictions severely limited his local job opportunities. His experience working through disruption explains the reason for relatively lower concern among workers. As he noted:

> "Initially, I was a bit concerned because most of the tasks that we used to do manually are now done by AI, especially writing. Most of it is done by AI. A client has a website, and they want to write content. So, most of these big things are done by AI. Initially, the workload went down, especially in those tasks. But what I have learned is, AI is still developing; I think it's still in the early stages. What happened is, most of the clients who tried using it, I think, realized there were still issues with quality. So, they still needed someone who is keen on writing to edit and maybe generate more accurate content than AI" (Theo, software and web developer).

Platform workers like Theo are often conditioned to adapt to shifting market dynamics, including technological disruptions. In the platform economy, adaptability is not just an edge but often a necessity, as workers face direct livelihood risks when they fail to adjust to changing client demands or emerging technology transitions, like AI. However, adaptation is a less risk-free process for firms. For clients, adopting AI too hastily carries its own set of risks, such as reduced quality, unforeseen costs, or workflow disruptions, especially when AI proves an imperfect substitute for human labour. This interplay of risks on both sides means that fluctuations in labour demand are a part of an ongoing recalibration process. As both workers and clients navigate these uncertainties, significant variation in the nature and volume of available work is likely before any semblance of stability emerges in the platform labour market. Amid this uncertainty, some workers go beyond reactive adjustment by actively anticipating market shifts and repositioning themselves accordingly, when their circumstances permit. Jasper, who transitioned from a solo platform freelancer to a successful entrepreneur and mentor by establishing his firm with a group of workers, explained:

> "The websites we used to work on, like three websites [in] a full month. But now, we can do three websites in a week. Traditionally, if someone wanted to do like an SEO copy for like, a landing page [of] a website, it's probably going to take you, like, three days before the project is done, but nowadays, with AI and just a few hours with proper prompt engineering[,] you have [it] done. I'm not really so worried about artificial intelligence, because it's here and we have to adapt to it. There's no way we can actually say it's not going to be part of us anymore, because [already] businesses, corporations, and even big companies, are actually using it. So, we also have to adapt to it and understand how we can actually include it in what we do as professionals" (Jasper, graphic designer, web developer, entrepreneur and trainer).

For professionals with established positions, such as Jasper and his team, recognizing emerging trends and staying ahead of ongoing labour market disruptions is essential to maintaining a competitive edge. Consequently, continuous upskilling, reskilling, and effective repositioning of their skills, combined with hard-earned credibility that attracts a steady stream of work, serve as vital safety nets against the disruptions. However, not all workers share this experience. For instance, Iris explained:

> "You might get to work today; tomorrow, you don't have work, and then you have bills here to pay. I see most of [the tasks] going to automation because most of [the clients], sometimes, they even tell you to test apps for them. Someone develops an app and wants it to be tested, like some here locally in Kenya. So, I think something [will change] in [the] future. It should be easier to test it, just wherever you are; you won't need someone in Kenya to do it for you. I think in [the] future, it won't be easy to work online. It concerns me a lot. Because I'm now thinking I have to think outside the box… what will I do if now we can't get to work online because maybe AI has replaced us and things like that?" (Iris, virtual assistant).

The volatile labour market and the transient nature of platform jobs can raise concerns about maintaining a consistent workload, particularly as automation and general-purpose AI increasingly compete with human tasks. For example, Iris turned to virtual assistance jobs through a platform after the very short three-month maternity leave made it challenging to return to her marketing job while caring for a newborn with health complications. In Kenya's rapidly developing economy, she seeks diverse opportunities where her local presence in Kenya offers her a competitive advantage. Still, the precarious nature of work availability, combined with the growing role of AI, remains a significant concern for her.

The contrasting experiences of Jasper, Iris, and many other workers demonstrate how AI-induced livelihood insecurities intensify when access to suitable work is already restricted, both on and off digital platforms. However, workers who are already in a position to maintain robust control and stability over their work prove more adept at navigating the disruptions driven by AI and the vulnerabilities it introduces. As a result, AI's impact on digital platform work is decidedly uneven. Its benefits or drawbacks hinge on each worker's unique livelihood context and are shaped by their ability to cultivate trust, leverage established profiles, and swiftly adapt to shifting market dynamics. In this rapidly evolving landscape, resilience and adaptability are not just assets but essential survival skills for sustaining a decent livelihood.

## The Gap Between AI Usage and AI Livelihood Impact Perception

While the interpretations above provide a plausible understanding of how early-stage empirical evidence on the impact of AI might be perceived, they also raise the question of what other factors could be at play if perceived job security is not always a significant determinant of concerns regarding AI-related job loss among workers. As illustrated in Figure 5.12, the empirical analysis of the relationship between levels of AI concern and AI usage shows that worry about AI is highest among those who use it the least.

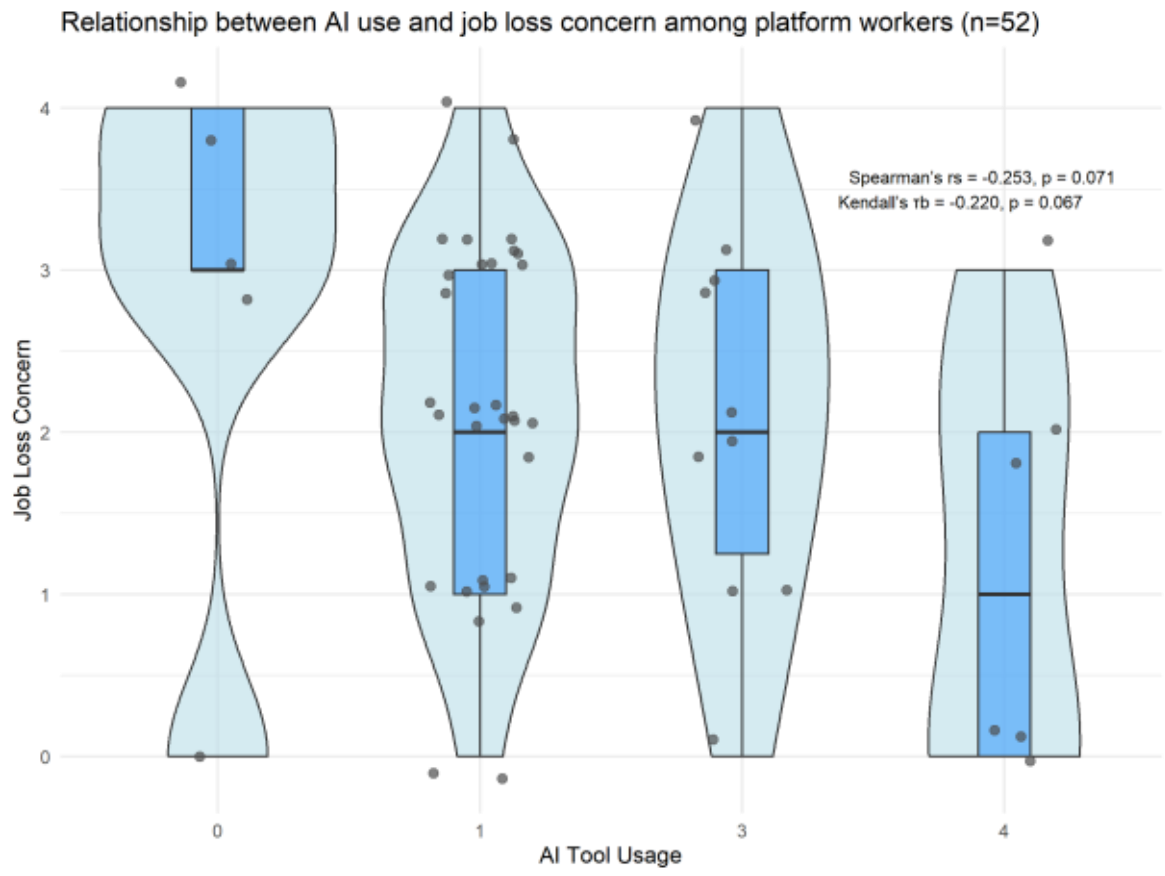

**Figure 12: AI Usage and Job Loss Concerns Among Kenyan Platform Workers[11]**

A likely explanation is that a lack of familiarity with AI and barriers to effectively utilizing AI tools to support workflow heighten uncertainty, exacerbating worker apprehension. These results, although somewhat statistically significant (p = .073), also indicate that usage alone accounts for only a small portion (6.3%) of this variation.

During the interviews, it became clear that although the narrative on AI often emphasizes productivity gains and potential job losses to automation, workers' views on the potential AI-driven work transformations were significantly shaped by their awareness that AI represents an advancement with the potential to create additional job opportunities for Kenyan workers. When asked how the expansion of AI is changing access to work, Marvin explained:

> “We have projects based on AI [,for example,] since last week, Freelancer is hiring a lot of Kenyans who are going to help AI determine the best responses based on languages. So, like, the Swahili language is mostly spoken in East Africa, [and] best African people have a higher chance of getting these jobs. So yes, we can be replaced on the high-paying skills and the low-paying ones for the AI; they are coming, but I feel we are going to be, at some point, replaced” (Marvin, web developer).

Marvin monitors emerging trends and opportunities as part of his livelihood strategy. Based on my interviews, Marvin and several other workers demonstrated greater awareness of AI’s impacts on work, as well as the nature of outsourced digital jobs in Kenya, which shapes their outlook.

Leading AI models have increasingly depended on freelance and BPO (business process outsourcing) workers from key outsourcing hubs such as Kenya, the Philippines, and India for critical tasks such as data labelling, content moderation, and linguistic annotation (M. Graham et al., 2017; Gray & Siddharth, 2019; Melia, 2020). These contributions are essential for optimizing AI performance, and for many platform workers, particularly in Kenya, the surge in AI-related work is seen as a significant opportunity, undermining any broader perception of AI as a livelihood threat. Interviews indicate that many workers strategically engage in short-term gigs (e.g., annotation, data labelling) as a gap-fill strategy, enabling them to secure income while exploring broader, more sustainable opportunities on digital platforms.

### The Delayed Onset of Automation and Substitution Impact

Linking empirical data on work demand and supply with the theorical projections of AI impact requires recognizing that any theoretical projections of automation and substitution assume highly efficacious AI tools and widespread AI adoption. That level of product reliability and adoption has not yet occurred and may not materialize to the extent predicted, depending on how firms and workers adapt to these technological changes and how regulations are implemented to mitigate adverse impacts on labour. The delayed onset of automation can also be attributed to the inherent flexibility of labour roles within firms, which enables companies to reassign workers to alternative tasks when specific roles become less relevant due to automation, thereby buffering the immediate displacement of workers. Sierra explained:

> “When we automated our customer service system, I was pushed to the back end. So, I do — I don't interact with customers like I used to.

[11] A simple linear regression was conducted to examine the relationship between AI tool usage and job loss concern. The model was marginally significant, $F(1, 50) = 3.35$, $p = .073$, explaining 6.3% of the variance ($R^2 = .063$). AI tool usage was negatively associated with job loss concern ($\beta = -0.24$, $SE = 0.13$, $t = -1.83$, $p = .073$), suggesting that higher AI tool usage is likely linked to lower job loss concern, though the effect was weak, likely due to the smaller sample group of this study.

> So, at the moment, I just, if a customer wants something changed, they create a ticket, [and] AI [creates] a ticket, it comes to us, and then we make the changes. The owners find an opportunity, or somewhere they can use you, because, as much as AI can do the front end and reply to customers automatically, that means that on the back-end side, where, if a customer has a request, someone needs to actually get the request done because the AI can only respond. Luckily for the company I work for, yes, there is [something] at risk, but they always have a way of trying to get something else for you if they decided to go the automated area" (Sierra, virtual assistant and customer support agent).

Sierra transitioned into virtual assistance work after feeling burned out from a local marketing job in Nairobi, but she found a long-term remote gig in customer service. Her perspective towards the livelihood risks of early-stage automation and labour substitution demonstrates how the flexibility of labour roles within firms can delay the immediate onset of automation. This reaffirms the position in existing literature that technological change often leads to a reallocation of labour rather than immediate replacement, as firms adjust by shifting workers to tasks that complement automation (Acemoglu & Restrepo, 2019). That said, another critical factor that must be considered is the relative cost of platform labour.

Many firms are drawn to a global workforce because it provides access to significantly lower-cost human labour (Graham & Anwar, 2019). This phenomenon mirrors the rapid expansion of the BPO industry in the past, which leveraged digitalization and task fragmentation to reduce costs by employing workers in developing regions. During the interview, while Sierra felt grateful that the American firm she worked for found her another role as the firm transitioned towards automation, she also showed her frustration towards the wage and benefits inequality that makes such work arrangements attractive for firms:

> "You know, they don't recognize local holidays. So, that's one thing. They don't, they don't also recognize the benefits of, probably generally, because in the companies that I work[ for], Americans get medical benefits. But remote workers do not; they get pension benefits that remote workers, they do not" (Sierra, virtual assistant and customer support agent).

Cost competitiveness often serves as a key rationale for moving away from traditional practices, such as offshoring work to BPO firms, hiring freelancers through digital platforms, or outsourcing tasks to AI. The decision to fully automate or substitute labour ultimately depends on the relative costs of human labour versus technology. Automation can yield significant cost savings and boost productivity in environments where labour is expensive. Conversely, when labour costs are low, the economic incentives for automation diminish, as the marginal efficiency gains may not justify the capital outlay and ongoing expenses (Autor, 2015; Howson et al., 2022).

In the case of AI-driven automation and labour substitution, cost-effectiveness is determined by a complex interplay of factors, such as acceptable level of AI advancements, capital availability, and labour market conditions. The result of this interplay is that the relative cost of platform labour is more competitive than traditional labour and, in many cases, AI-automation. While such automation might promise higher productivity, its adoption is currently curtailed in low-wage settings where the marginal gains do not outweigh the initial costs and associated AI risks for firms. Consequently, the effects of automation and substitution are unlikely to be immediately revealed in their true form in many segments of the gig economy while the relative cost of platform labour remains low, and if AI adoption remains low among firms due to a range of socio-technical reasons, including the error-prone nature of AI tools[12].

### The Interplay of Paid and Unpaid Work

One of the key premises of this research is recognizing that AI tools affect the unpaid components of freelance work, potentially transforming conditions for workers, but only if AI adoption, accessibility, and user proficiency remain largely frictionless. Unpaid work accounts for a significant portion of human productive effort, often comparable in duration and impact to paid work. Unpaid work is an inherent feature of platform-based freelancing, functioning as a hidden cost necessary to secure sufficient paid opportunities while simultaneously reducing the time available to complete remunerated tasks. In the context of the platform economy, this invisibility is particularly problematic, as many gig workers invest significant time in auxiliary tasks.

[12] A pressing barrier to the utility of generative AI is that the predicted text output often appears authentic and factual at the surface level, yet its fabricated or falsified nature only becomes evident upon closer examination. This phenomenon, often referred to as AI hallucinations, indicate the useability barriers. Such hallucinations can lead to misleading information, underscoring the automation risks for firms and the continued need for expert oversight. For more on this, see (Emsley, 2023; Salvagno et al., 2023). Also see (Gosmar & Dahl, 2025) on how recent advances have demonstrated promising strategies to mitigate these errors by refining and clarifying speculative content, yet significant challenges remain.

In the field research survey, I asked workers which aspects of their jobs could be made more efficient with AI-based tools. A significant proportion of responses pointed to the perceived role of AI tools in transforming unpaid work. Roughly 10% of responses emphasized using AI tools solely for job application processes, such as drafting proposals, writing cover letters, and searching for suitable gigs, which this research classifies as unpaid tasks despite requiring considerable time and energy. Additionally, about 48% of participants highlighted AI's role in supporting activities such as extensive research, idea generation, and client communications. While these functions are crucial for securing paid gigs, they often remain auxiliary and largely uncompensated, underscoring the hidden labour costs inherent in the freelance platform economy.

Could AI reduce the burden of unpaid work for freelance workers and help them overcome the autonomy paradox? [13] The significantly high automation, substitution, and enhancement scores for unpaid tasks across various platform work categories indicate that a widespread adoption of AI is likely to substantially reduce a large amount of unpaid work in all categories. It should not be overlooked that unpaid work also requires considerable creativity and a personal touch, and it can often set a freelancer apart from the rest in a saturated labour market. Unpaid work in the platform economy also includes emotional labour, the strategic personalization of interactions used by workers to cultivate client relationships necessary for building long-term career growth opportunities (Alacovska et al., 2024). While AI's capacity to decrease unpaid work could provide substantial support, it also creates a new dilemma. The reduction in unpaid work relies on the same extensive adoption of AI in the workplace, which poses a risk of labour substitution and automation in paid positions.

I refer to this phenomenon as the enabling-obsolescence paradox, where the automation of both paid and unpaid tasks mutually reinforces each other. More importantly, it also produces the composite consequences of unpaid labour automation, resulting in the collective dehumanization of work, irrespective of its paid or unpaid status. By using the term enabling-obsolescence paradox, I emphasize how, in the early stages, AI support can reduce the tediousness of writing numerous job proposals to secure limited work or assist with the error-prone aspects of managing client communication in globalized work environments. However, these efficiency gains come at a significant cost of reducing the already limited human element of worker-work provider interactions that underpin organic interpersonal interactions, synergistic creativity, and authentic workplace engagement. Suppose the significance of such meaningful professional relationships diminishes; it will be a substantial leap towards replacing paid human labour with AI capital, and platforms are at the forefront of this change.

My position is that the drive for the deeper integration of AI in freelance work is inherently connected to the platformization of work and the business model of platforms that promote hasty technology acceptance (Silva Neto, 2019). Unsurprisingly, digital labour marketplaces are actively reshaping freelance work through this deeper integration of AI [14]. These platforms promote and integrate AI utility for freelancers, providing them with helpful tools for drafting proposals, collaboration with clients, and streamlining communication that not only supports workers but also redefines workplace norms.

[13] The autonomy paradox refers to the contradiction in which digital platform workers, attracted by promises of flexibility and the ability to be their own boss, ultimately discover that they must work long, irregular, and unsocial hours. Instead of enjoying true autonomy over their schedules, these workers are compelled to remain constantly available to secure work and maintain high ratings on the platform. This situation traps them in a cycle where their "free" time is effectively consumed by unpaid labour, undermining the very freedom they sought. For more on this, see (Alacovska et al., 2024; Mazmanian et al., 2013; Pulignano et al., 2024)

[14] For instance, Upwork, one of the largest freelancing platforms, is evolving from a destination for on-demand hiring of human labour to one for on-demand work outcomes. This change is subtle but important. The platform offers integration support for a range of third-party AI tools, as well as its own AI model, called "Uma". On the workers' end, these features are expected to assist with tasks such as drafting proposals, brainstorming, data management

However, by normalizing the broader use of these tools, the platforms condition the labour market to a broader acceptance of AI tools (even if both sides of the labour market recognize the involvement of AI tools), gradually accepting dehumanized interactions[15] and work processes as the new normal and replacing the nuanced and empathetic aspects of interaction with automated processes. Simultaneously, these platforms extend AI-driven solutions to work providers, transforming the traditional mediation between human labour and clients into one where AI agents increasingly assume the roles held by human workers and ultimately pave the way for AI capital to supplement human labour.

### The Forms of Early-Stage Automation and Substitution in the Platform Economy

While empirical observations reveal workers' perceptions of AI's impact, a critical gap remains in identifying and understanding the early-stage shifts in the job market that signal the onset of automation and task substitution. When it comes to reshaping existing tasks, a more plausible empirical scenario is that the initial phases of automation and substitution would manifest instead as task enhancement, benefiting workers more before the labour market adjusts to these transformations.

In the early stages, the AI-driven tools may improve productivity, increasing output without immediately displacing workers. However, as market pricing mechanisms and the skill premium evolve in response to AI adoption, the long-term implications for job security and income distribution may become more pronounced. Under this scenario, work that is likely to see higher automation and substitution under peak-AI would experience early adoption and increased utility of these tools, gradually reducing the skill premium for tasks that impact the livelihood of workers if they cannot transition to other, higher-paying gigs.

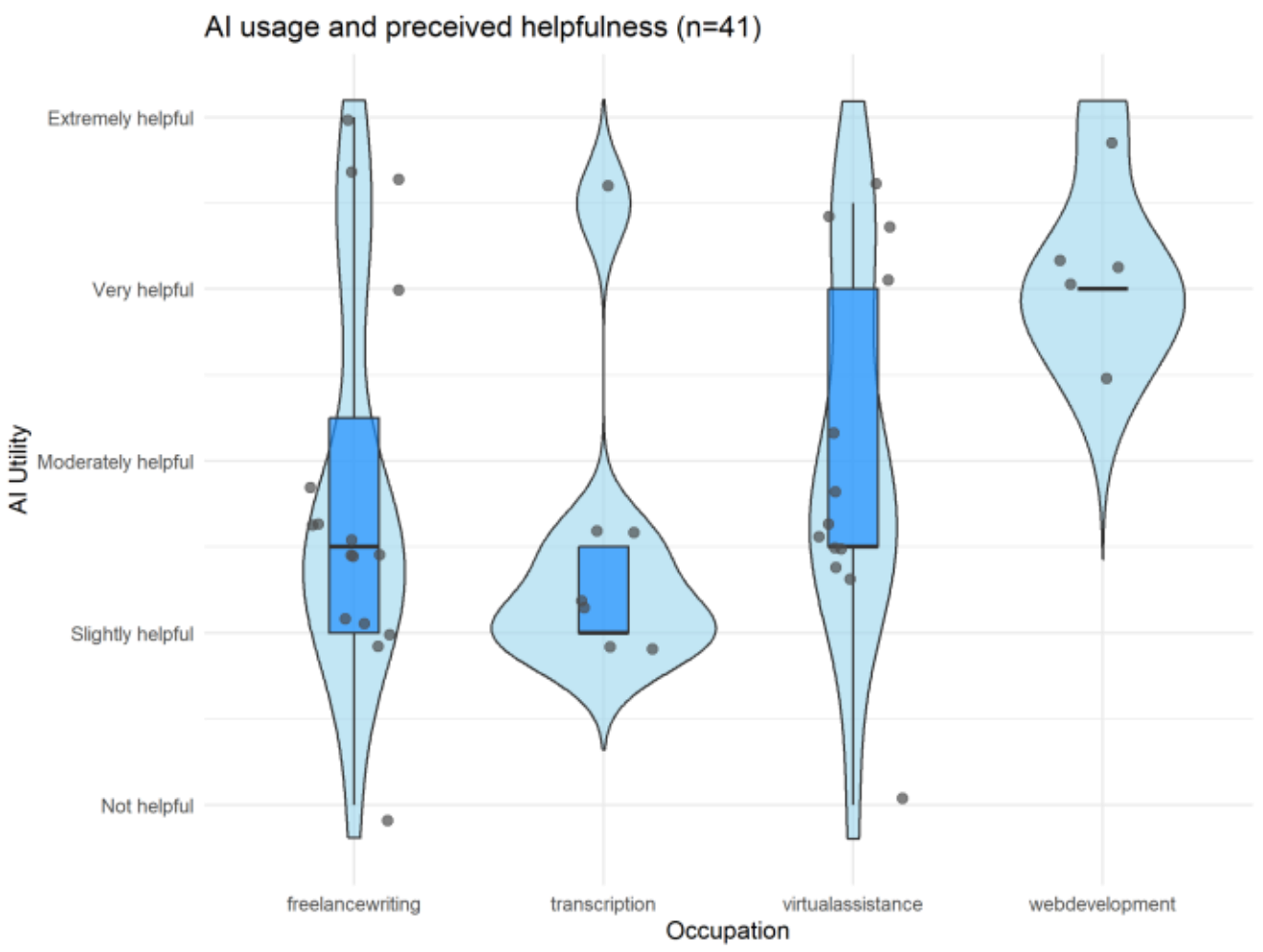

**Figure 13: Perception of AI Utility Among Kenyan Platform Workers**

This delayed automation impact hypothesis is grounded in empirical data presented in Figure 5.13 and the theoretical analysis results presented in the previous section. It indicates that, regarding AI helpfulness, workers involved in work categories with a higher proportion of automatable tasks that lead to greater livelihood vulnerability, such as those in software and web development gig work (Figure 5.14), are more likely to perceive AI tools as helpful. In contrast, freelance writers (Figure 5.15) are less likely to experience a similar long-term impact involving automation, substitution, and vulnerability.

---

and streamlining communication. Concurrently, Upwork now offers clients fully managed AI services that include the creation of custom AI agents tailored to handle a variety of tasks. (Upwork Global Inc., n.d). This development not only streamlines project management and outcome delivery but also signifies a shift in platform mediation of human labour to increasingly transitioning toward mediation of AI agents.

[15] In the broader context of technological substitution, work dehumanization is a threat to employee–employer relationships. This concern is twofold: first, the transition to algorithmic processes erodes the nuanced human presence fundamental to authentic interpersonal interactions; second, the values inherent in human judgment often diverge from the vague metrics embedded in algorithms. See (Fritts & Cabrera, 2021) on how the use of recruitment algorithms results in such dehumanization.

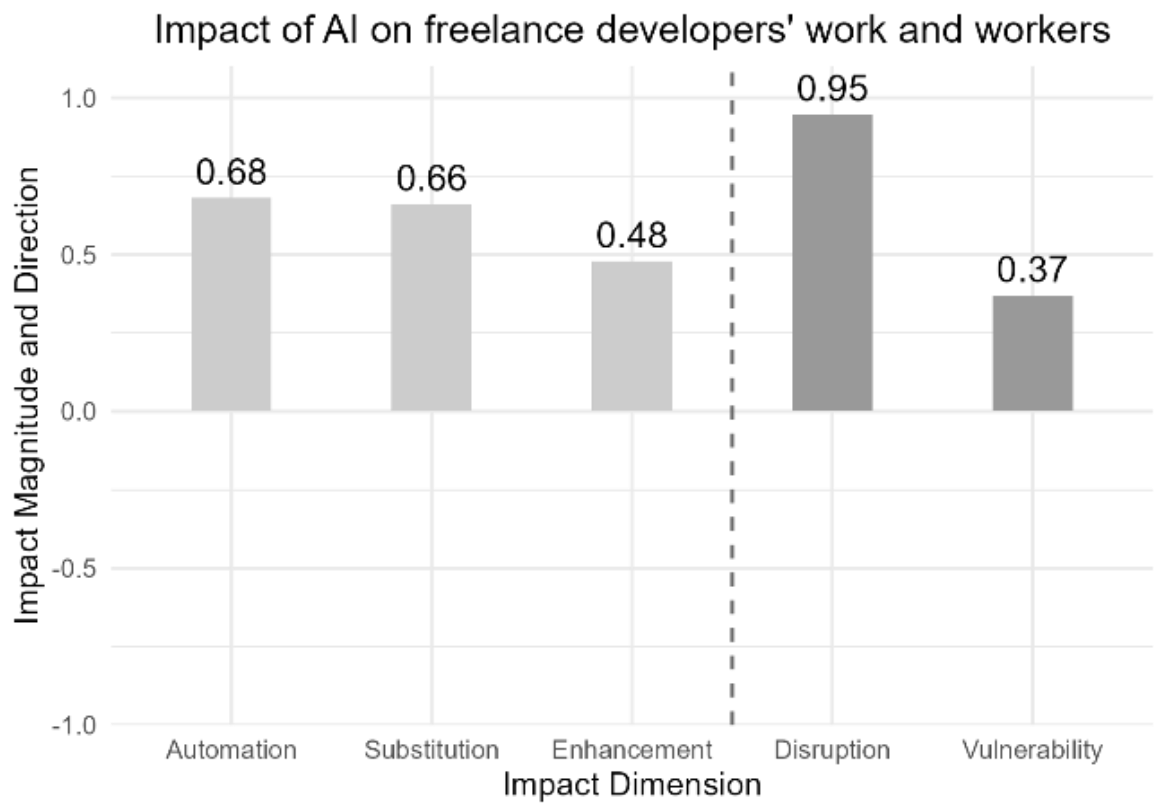

**Figure 14: Impact of AI on Freelance Web and Software Development Work**

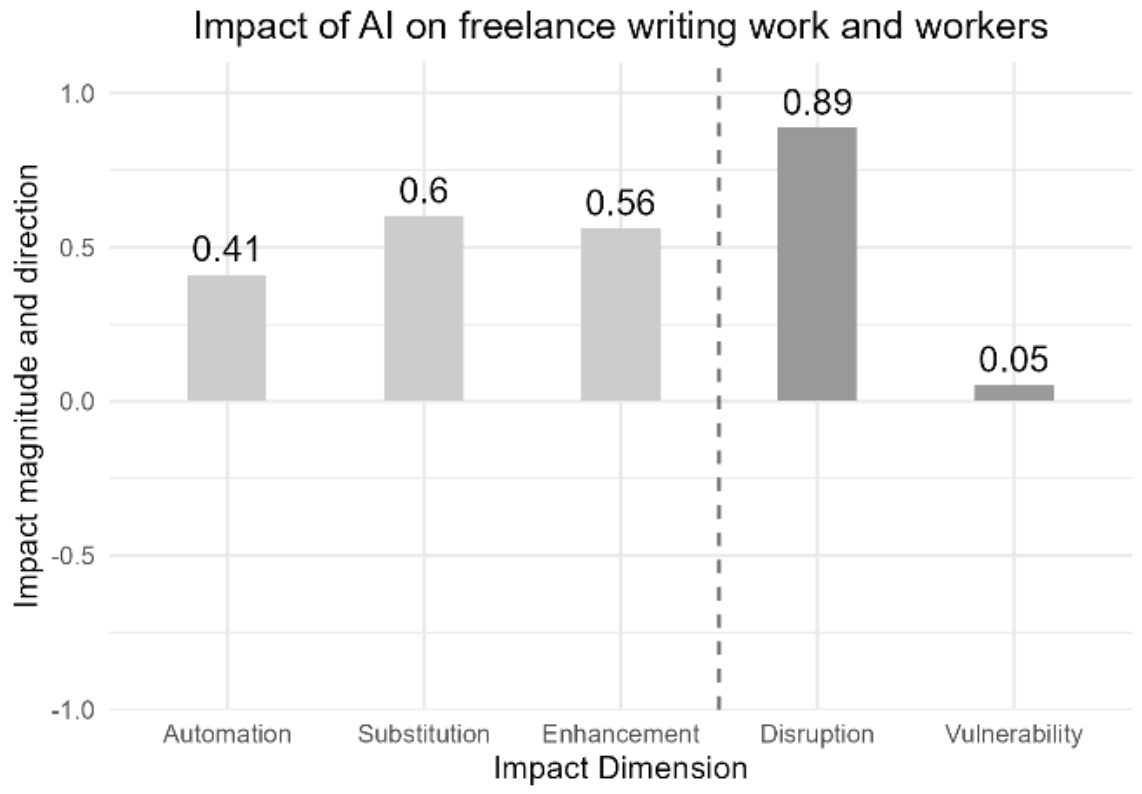

**Figure 15: Impact of AI on Writing Jobs**

The interview data also provides support for this argument, suggesting that when access to the amount of work is not a concern, the perspective of tech workers such as Jasper and Theo revealed how the benefits of AI have a considerable positive impact on their work:

> "For me, AI is a development in the right place, because now things [tasks] that we used to do, we used to take a long time to actually do. We can actually automate very easily and have them done within, like, maybe [a] few days" (Jasper, graphic designer and web developer).

> "I feel like AI is like making everything easier"[16] (Theo, software and web developer).

Jasper and Theo express a utilitarian perspective, where the immediate benefits of efficiency and ease of work outweigh concerns about the long-term sustainability of short-term, often transitory, platform-based tasks. When the focus is on productivity and efficiency, workers tend to assess AI in a largely positive light, even if they are aware of potential long-term risks. However, some workers exhibited a notable indifference toward these technological transitions. For them, while automation and substitution may compromise the attractiveness or availability of certain types of work, these changes do not pose fatal threats to their overall livelihood. The consistently high disruption scores across all work categories further illuminate this perspective. For example, Mable and Clara explained:

> "I'm not concerned because now, if I find out that it [AI] is replacing some aspect, I go on to the next thing and look for something that is going to help me move further and become better" (Mable, Freelance writer and accountant).

> "For me, it is not a big challenge; it is just about being open-minded. Things are going to change always, just have that in mind. They're gonna change either in your way or in another person's way, not necessarily the way you like it or the way you want it. That acceptance is what makes you move forward" (Clara, freelance writer).

A notable observation from the theoretical measurement of AI impact presented in the previous section is the exceptionally high level of disruption, with scores ranging from 0.89 to 0.95 on a scale of 0 to 1. Such elevated disruption levels reflect that, regardless of whether AI's impact manifests as substitution, enhancement, or automation, workers must proactively adopt strategies to remain competitive in an increasingly dynamic labour market.

To empirically contextualize this high level of platform work disruption, and Clara and Mable's comments, it is worth noting that Kenyan platform workers have historically operated within a gig economy characterized by inherent instability, where technological disruptions extend beyond recent advancements in AI technology (Melia, 2020). Although platform work could be segmented into various categories to understand the magnitude of such technology's impact on platform work, the transitory nature of platform work persuades workers to remain vigilant, continually seeking new niches and switching work categories when their current gigs become less lucrative. This precariousness conditions workers to view work categories as temporary, driving them to adapt swiftly and cultivate resilience in the face of constant disruption and livelihood risks. Ultimately, the fluid landscape of platform work compels Kenyan workers to adopt an opportunistic mindset, enabling them to worry less and navigate disruptions effectively, continually reshaping their career trajectories in response to evolving market demands.

[16] Theo shared this perspective during our interview in March 2024. By the end of September 2024, he contacted me to express that he was struggling to secure similar software and web development gigs and was considering a transition into video editing and marketing.

### Navigating AI Disruption Through Upskilling and Adaptability

Platform workers are acutely aware that the evolving nature of their tasks necessitates attentiveness to changing market dynamics and the need for reskilling. In my survey, over 92% of participants strongly agreed that acquiring AI-related skills is essential for maintaining job security in the face of potential job displacement due to AI. In the survey, participants were asked an open-ended question about how aspects of their work might be replaced by AI-based tools and the support or training required to navigate this transition. Responses primarily highlighted the need for a better understanding of AI integration into their workflows, emphasizing the development of skills as a priority in an evolving job market.

A semantic analysis of these responses revealed that approximately 60% of participants emphasized specialized technical skills, such as prompt engineering, programming, and data analytics, as crucial for preparing to work with AI. Roughly half of the responses focused on practical AI integration, indicating a desire for concrete guidance on incorporating AI into writing, design, and planning tasks. About 40% highlight the need for ongoing, technology-oriented learning and accessible resources, often referencing cost concerns and the importance of formal or informal programs to adapt effectively to AI-based tools. A smaller portion, around 20%, mentioned business or soft skills training, including marketing and creative thinking, as equally vital in staying relevant alongside AI. Finally, a subset of around 10% explicitly referenced ethical and societal considerations, underscoring the importance of responsible AI usage and professional standards for freelancers. These proportions illustrate that while the majority prioritize direct technical training to utilize AI tools, there is also notable interest in broader skill development, continuous education, and ethical awareness in an AI-driven landscape.

Evidently, platform workers in Kenya demonstrate remarkable resilience in navigating evolving labour market dynamics, particularly in response to technological transformations. While concerns about AI-driven displacement are widespread, this does not necessarily indicate an adaptability crisis; instead, it reflects workers' awareness of the changes reshaping digital labour. Evidence from the survey (Table 5.8) shows that a substantial majority, 78.85%, express confidence in their ability to transition to new job opportunities if AI-based tools were to replace some aspects of their current work. In contrast, 17.31% remain uncertain, and only 3.85% anticipate significant challenges to adaptability.

**Table 0.8**
**Perceived Adaptability of Skills to New Job Opportunities Amid AI Disruption**

| | Frequency | Percentage | Cumulative |
|---|---|---|---|
| Probably yes | 23 | 44.23 | 44.23 |
| Definitely yes | 18 | 34.62 | 78.85 |
| Might or might not | 9 | 17.31 | 96.15 |
| Definitely not \| | 2 | 3.85 | 100.00 |
| Total | 52 | 100.00 | |

This elevated level of confidence aligns with findings from digital labour research, which indicate that platform workers consistently engage in upskilling, diversifying their tasks, and strategically navigating their careers to maintain employability in an unpredictable market (Elbanna & Idowu, 2021; Mark Graham et al., 2017; Melia, 2020). Unlike workers in traditional employment structures, freelancers operating in the digital gig economy, particularly in countries such as Kenya, are conditioned to embrace adaptability as a core survival strategy. They often juggle multiple tasks across different skill levels and industries to hedge against market volatility. As evident from the interviews, several workers shared this sentiment:

> “One of my strengths is adaptability. If AI starts taking over, I’d want to learn how to use it and find my place within that system. So, yes, it might take my job, but [it] will not take my ability to take on other jobs by using it” (Simon, freelance transcriptionist).

> “Already, like for me, [we're] already getting into machine learning, data science, [and] also getting into artificial intelligence itself, because it's already here. And for us to say, to stay sustainable and to also stay in business, we also have to adapt. So, for me, I force everyone around me to get started with artificial

> intelligence as early as now" (Jasper, freelance web developer and graphic designer).

> "Quite concerned, but maybe not worried, because I have also integrated AI into my workflow. [My] thinking is that freelancing is a one-person business, so we are actively including aspects of AI which we think would help the beneficiary" (Owen, freelance transcriptionist and trainer).

Concern and confidence coexist among experienced workers like Simon, Jasper and Owen. Their concern stems from a clear understanding of the potential disruptions introduced by AI, which necessitate continuous upskilling, diversification of services, and exploration of new opportunities emerging from technological change. At the same time, confidence arises from their adaptive orientation. Workers are structurally conditioned to embrace adaptability not just as a strategy but as a fundamental requirement for survival and advancement in an uncertain and rapidly evolving labour market.

The significance of this adaptability is evident in many workers, such as Owen, who has worked in the transcription services field since 2007. Reflecting on his early career, Owen described transcribing voicemails and converting them into written text for clients in the United States, a clear example of how hidden human labour has always underpinned seemingly automated digital processes for decades, and workers continue to cultivate transition pathways. Today, alongside his continued roles as a transcriptionist and virtual assistant, Owen's primary job is to provide skills training to refugees in Dadaab and Kakuma, equipping them with the diverse skills needed to succeed on freelance platforms. His optimism about adaptability and managing technological transition extends to his trainees, as he explains:

> "They [workers] have the fortitude to persevere and innovate. Some skills, especially language-related skills like Dadaab and Somali, are protected niches. With adequate support and mentorship, they [workers] can transition" (Owen, freelance transcriptionist and trainer).

Owen's cautiously optimistic response illustrates not only Kenyan platform workers' inherent capacity to adapt but also their perceived need to do so in order to sustain their livelihoods. This dual dynamic helps explain why many view AI-driven disruptions less as existential threats and more as manageable challenges within the evolving platform economy. These perspectives also reveal a critical distinction between the vulnerability of individual workers' livelihoods and the broader vulnerability associated with specific categories of platform work. While specific types of platform tasks and work may be more immediately vulnerable to AI-driven disruption, similar technological shocks are reshaping work globally by fragmenting traditional permanent roles into discrete, task-based units. Some of these tasks might be automated, while others could shift to different labour structures, including digital platforms, and Kenyan freelancers seem ready to adapt and strategically position themselves within this changing landscape.

## *Conclusions*

This research emphasizes the adoption of a nuanced, context-sensitive approach to understanding and managing AI's impact on digital platform labour. By bridging theoretical insights with empirical realities, the study contributes to contemporary debates on technological transformation, worker vulnerability, and adaptive strategies within the rapidly evolving digital economy. It discusses the gaps between theoretical and empirical understanding of AI's impact on work. The task-based analytical framework employed in this research offers a robust methodological approach to identifying which occupational tasks are more susceptible to automation and which may effectively resist it. This framework presents a replicable model for future studies exploring the labour impacts of AI in diverse contexts while leveraging novel research technologies, such as using LLM models to support analyses.

The AI impact scores derived from the presented model indicate substantial occupational changes driven by AI, highlighting high automation risks in coding tasks, significant substitution effects in transcription services, and notable enhancements in virtual assistance roles. These findings underscore AI's dual role in simultaneously augmenting and potentially displacing human labour, reinforcing the complexity of its impact, which is heavily mediated by the intrinsic characteristics of specific tasks.

Empirical evidence from surveys and interviews with 52 Kenyan freelancers partly validates the theoretical predictions. However, these projections remain inherently probabilistic, assuming frictionless and extensive AI adoption, a scenario not fully realized in empirical contexts and an assumption that may not materialize uniformly. Still, the empirical section of the study highlights concrete shifts in freelancers' job tasks, increased reliance on AI tools for paid and unpaid tasks, and varied concerns about future job security and adaptability. This alignment particularly emphasizes the resilience and adaptive capacity of Kenyan platform workers, who consistently perceive technological disruptions as manageable challenges rather than existential threats. Such resilience is not mere optimism but is deeply rooted in structural realities of platform workers, including limited institutional safety nets and continuous exposure to job volatility, which have historically conditioned freelancers to develop proactive skills, be flexible, and engage in continuous learning.

Despite strong findings in this article, several limitations remain. Notably, disruption vulnerability formulas depend on multiplicative models and Gaussian thresholds, where hyperparameters are determined by subjective judgment. A similar replication limitation affects the novel research methods that utilize LLM-based scoring against task characteristics, such as task modularity and social complexity. Additionally, a sole focus on specific tasks may misjudge broader political-economic factors and underestimate systemic drivers of AI adoption. Future research should broaden its geographic scope, include longitudinal analyses to track the evolving impacts of AI, and consider client-side dynamics.